%% file: doc.tex
\PassOptionsToPackage{dvipsnames,svgnames}{xcolor}

\documentclass[acmsmall,screen]{acmart}

\setcopyright{none}
\renewcommand\footnotetextcopyrightpermission[1]{}
\usepackage{subcaption}
\usepackage{tabulary}
\usepackage{threeparttable}
\usepackage{booktabs}
\usepackage{relsize,xspace}
\usepackage{multirow}
\usepackage{enumitem}
\usepackage{balance}
\usepackage{array}
\usepackage{hyperref}
\usepackage{multicol}
\usepackage{microtype}
\usepackage{makecell}
\usepackage{graphicx}
\usepackage[textsize=tiny]{todonotes}
\usepackage{listings}
\usepackage[table,xcdraw]{xcolor}
\usepackage[normalem]{ulem}
\usepackage[capitalize]{cleveref}
\Crefname{lstlisting}{Listing}{Listings}
\crefname{lstlisting}{listing}{listings}
\useunder{\uline}{\ul}{}

\input{Macros.tex}

\input{generated_tangling_macros.tex}

\newcommand\parhead[1]{\vspace{.5mm}\noindent\textbf{{#1}}}

\newcommand{\quotebox}[3]{\vspace{.5em}\noindent\begin{tikzpicture}
    \node[align=center,draw,thin,minimum width=\columnwidth,inner sep=2.2mm] (titlebox)%
    {\parbox{0.95\columnwidth}{\looseness=-1\noindent\textit{#2\vspace{0.1cm}}}};
    \node[label=left:{\colorbox{white}{\small #1}}] (W) at (titlebox.south east) {};%
    \end{tikzpicture}\vspace{-15pt}}

\begin{document}
\fancypagestyle{firstpagestyle}{%
  \fancyhf{}%
  \renewcommand{\headrulewidth}{0pt}%
  \renewcommand{\footrulewidth}{0pt}%
  \fancyfoot[RO,LE]{\small\thepage}%
}
\fancypagestyle{standardpagestyle}{%
  \fancyhf{}%
  \renewcommand{\headrulewidth}{0pt}%
  \renewcommand{\footrulewidth}{0pt}%
  \fancyhead[LE]{\thepage}%
  \fancyhead[RO]{\thepage}%
  \fancyhead[RE]{Hermann et al.}%
  \fancyhead[LO]{\shorttitle}%
}
\pagestyle{standardpagestyle}

\newcommand{\tb}[1]{{\textsf{\textbf{TB}}[\smaller\sffamily\color{orange} #1]}}
\newcommand{\svp}[1]{{\textsf{\textbf{SP}}[\smaller\sffamily\color{ForestGreen} #1]}}
\newcommand{\kh}[1]{{\textsf{\textbf{KH}}[\smaller\sffamily\color{blue} #1]}}
\newcommand{\rwi}[1]{{\textsf{\textbf{RW1}}[\smaller\sffamily\color{red} #1]}}
\newcommand{\rwii}[1]{{\textsf{\textbf{RW2}}[\smaller\sffamily\color{red} #1]}}
\newcommand{\rwiii}[1]{{\textsf{\textbf{RW3}}[\smaller\sffamily\color{red} #1]}}

\newcommand{\sumbox}[1]{
\noindent\fbox{\parbox{\dimexpr \linewidth-2\fboxsep-2\fboxrule}{\noindent
		\textit{#1}
	}}\smallskip
}

\title{Security Is More Than a Library Call: How Security Features Live in Code}
\settopmatter{authorsperrow=4}
\author{Kevin Hermann}
\orcid{0009-0004-6207-4045}
\affiliation{%
	\institution{Ruhr University Bochum}
	\country{Germany}
	\city{Bochum}
}

\author{Sven Peldszus}
\orcid{0000-0002-2604-0487}
\affiliation{%
	\institution{Chalmers\,|\,University of Gothenburg}
	\country{Sweden}
	\city{Gothenburg}
}

\author{Thorsten Berger}
\orcid{0000-0002-3870-5167}
\affiliation{
	\institution{Ruhr University Bochum and Chalmers\,|\,University of Gothenburg}
	\city{Bochum}
	\country{Germany and Sweden}
}

%
%

%

\begin{abstract}
Implementing security features---functionalities that protect sensitive data or prevent malicious actions by attackers---is important for ensuring the security and integrity of software systems.
Correctly implementing access control, cryptography, or other security features is challenging as they require substantial domain expertise, careful design, and custom implementation to integrate them within the software system.

Previous work has thoroughly studied security awareness, perception, practices, and expertise via surveys, interviews, and experiments.
While they have shown that the implementation of security features are often supported by security libraries and frameworks, they have also shown that developers rarely are security experts, and make mistakes that introduce vulnerabilities into software when using them.
Despite this extensive knowledge of security practices and developer behavior, we still lack a comprehensive understanding of how security features actually manifest at the code level across full software systems.

We close this gap by conducting a mining study of security features in 9 popular and large open-source Java systems such as Hadoop, Signal Server, and Tomcat.
We manually inspected 2,127,761 lines of code across 19,121 files, identifying and annotating 183,395 lines implementing 561 security features corresponding to 54 security features in our taxonomy. 
We analyzed the characteristics of the identified security features, such as their size, scattering, tangling, common implementation patterns, and the use of internal and external functionalities.
Our findings show that security features are far more than mere calls to external libraries.
Security features, such as access control, can grow large in size, scatter across the whole codebase, and frequently tangle with each other.
External security libraries are commonly used, but they require substantial amounts of code for integration.
Once integrated, developers reuse their own functionality extensively throughout the codebase.
We publish our annotated dataset for future research on security features in software systems.
\end{abstract}

\maketitle

\input{01_Introduction.tex}

\input{02_Background.tex}

\input{03_Methodology.tex}

\input{04_Dataset.tex}

\input{05_Results.tex}

\input{06_Discussion.tex}

\input{07_Limitations.tex}

\input{08_Conclusion.tex}






\section*{Data Availability}
We make our annotated dataset and analysis scripts available\,\cite{securityfeatureminingdataset}.
Upon acceptance of this manuscript, the repository will be made publicly accessible on Zenodo.


\bibliographystyle{IEEEtran}
\bibliography{IEEEabrv,references}

\end{document}

%% file: Macros.tex
\newcommand{\accesscontrolLLocPctAvg}{6.76}
\newcommand{\accesscontrolLLocPctStd}{7.31}
\newcommand{\accesscontrolScatMin}{35}

\newcommand{\accesscontrolScatMedian}{151}

\newcommand{\accesscontrolScatMax}{1,497}
\newcommand{\accesscontrolScatAvg}{301}
\newcommand{\accesscontrolScatStd}{437}
\newcommand{\accesscontrolMScatMin}{85}

\newcommand{\accesscontrolMScatMedian}{857}

\newcommand{\accesscontrolMScatMax}{8,165}
\newcommand{\accesscontrolMScatAvg}{1,558}
\newcommand{\accesscontrolMScatStd}{2,405}
\newcommand{\accesscontrolScatPctMin}{2.12}

\newcommand{\accesscontrolScatPctMedian}{12.14}

\newcommand{\accesscontrolScatPctMax}{39.79}
\newcommand{\accesscontrolScatPctAvg}{13.87}
\newcommand{\accesscontrolScatPctStd}{10.76}
\newcommand{\accesscontrolMScatPctMin}{0.39}

\newcommand{\accesscontrolMScatPctMedian}{5.32}

\newcommand{\accesscontrolMScatPctMax}{32.34}
\newcommand{\accesscontrolMScatPctAvg}{8.12}
\newcommand{\accesscontrolMScatPctStd}{9.02}

\newcommand{\accesscontrolLocSizeMax}{1,439.0}

\newcommand{\cryptographyLLocPctAvg}{3.32}
\newcommand{\cryptographyLLocPctStd}{3.96}
\newcommand{\cryptographyScatMin}{4}

\newcommand{\cryptographyScatMedian}{42}

\newcommand{\cryptographyScatMax}{584}
\newcommand{\cryptographyScatAvg}{132}
\newcommand{\cryptographyScatStd}{171}
\newcommand{\cryptographyMScatMin}{13}

\newcommand{\cryptographyMScatMedian}{133}

\newcommand{\cryptographyMScatMax}{2,920}
\newcommand{\cryptographyMScatAvg}{713}
\newcommand{\cryptographyMScatStd}{1,037}
\newcommand{\cryptographyScatPctMin}{0.19}

\newcommand{\cryptographyScatPctMedian}{7.42}

\newcommand{\cryptographyScatPctMax}{20.29}
\newcommand{\cryptographyScatPctAvg}{7.09}
\newcommand{\cryptographyScatPctStd}{6.04}
\newcommand{\cryptographyMScatPctMin}{0.06}

\newcommand{\cryptographyMScatPctMedian}{1.76}

\newcommand{\cryptographyMScatPctMax}{10.10}
\newcommand{\cryptographyMScatPctAvg}{3.17}
\newcommand{\cryptographyMScatPctStd}{3.26}

\newcommand{\cryptographyLocSizeQThree}{4.0}
\newcommand{\cryptographyLocSizeMax}{5,213.0}

\newcommand{\securitymonitoringLLocPctAvg}{0.27}
\newcommand{\securitymonitoringLLocPctStd}{0.48}
\newcommand{\securitymonitoringScatMin}{2}

\newcommand{\securitymonitoringScatMedian}{14}

\newcommand{\securitymonitoringScatMax}{311}
\newcommand{\securitymonitoringScatAvg}{51}
\newcommand{\securitymonitoringScatStd}{94}
\newcommand{\securitymonitoringMScatMin}{4}

\newcommand{\securitymonitoringMScatMedian}{16}

\newcommand{\securitymonitoringMScatMax}{902}
\newcommand{\securitymonitoringMScatAvg}{214}
\newcommand{\securitymonitoringMScatStd}{361}
\newcommand{\securitymonitoringScatPctMin}{0.19}

\newcommand{\securitymonitoringScatPctMedian}{1.17}

\newcommand{\securitymonitoringScatPctMax}{4.65}
\newcommand{\securitymonitoringScatPctAvg}{1.70}
\newcommand{\securitymonitoringScatPctStd}{1.48}
\newcommand{\securitymonitoringMScatPctMin}{0.02}

\newcommand{\securitymonitoringMScatPctMedian}{0.23}

\newcommand{\securitymonitoringMScatPctMax}{2.76}
\newcommand{\securitymonitoringMScatPctAvg}{0.62}
\newcommand{\securitymonitoringMScatPctStd}{0.82}

\newcommand{\securitymonitoringLocSizeQThree}{2.0}
\newcommand{\securitymonitoringLocSizeMax}{1,222.0}

\newcommand{\securedatahandlingLLocPctAvg}{1.49}
\newcommand{\securedatahandlingLLocPctStd}{3.03}
\newcommand{\securedatahandlingScatMin}{2}

\newcommand{\securedatahandlingScatMedian}{32}

\newcommand{\securedatahandlingScatMax}{228}
\newcommand{\securedatahandlingScatAvg}{55}
\newcommand{\securedatahandlingScatStd}{69}
\newcommand{\securedatahandlingMScatMin}{4}

\newcommand{\securedatahandlingMScatMedian}{142}

\newcommand{\securedatahandlingMScatMax}{853}
\newcommand{\securedatahandlingMScatAvg}{205}
\newcommand{\securedatahandlingMScatStd}{261}
\newcommand{\securedatahandlingScatPctMin}{0.09}

\newcommand{\securedatahandlingScatPctMedian}{2.91}

\newcommand{\securedatahandlingScatPctMax}{11.99}
\newcommand{\securedatahandlingScatPctAvg}{3.49}
\newcommand{\securedatahandlingScatPctStd}{3.75}
\newcommand{\securedatahandlingMScatPctMin}{0.01}

\newcommand{\securedatahandlingMScatPctMedian}{0.70}

\newcommand{\securedatahandlingMScatPctMax}{10.22}
\newcommand{\securedatahandlingMScatPctAvg}{1.78}
\newcommand{\securedatahandlingMScatPctStd}{3.07}

\newcommand{\securedatahandlingLocSizeQThree}{3.0}
\newcommand{\securedatahandlingLocSizeMax}{740.0}

\newcommand{\systemstateprotectionLLocPctAvg}{1.05}
\newcommand{\systemstateprotectionLLocPctStd}{1.55}
\newcommand{\systemstateprotectionScatMin}{5}

\newcommand{\systemstateprotectionScatMedian}{22}

\newcommand{\systemstateprotectionScatMax}{119}
\newcommand{\systemstateprotectionScatAvg}{47}
\newcommand{\systemstateprotectionScatStd}{45}
\newcommand{\systemstateprotectionMScatMin}{8}

\newcommand{\systemstateprotectionMScatMedian}{52}

\newcommand{\systemstateprotectionMScatMax}{724}
\newcommand{\systemstateprotectionMScatAvg}{178}
\newcommand{\systemstateprotectionMScatStd}{228}
\newcommand{\systemstateprotectionScatPctMin}{0.24}

\newcommand{\systemstateprotectionScatPctMedian}{1.81}

\newcommand{\systemstateprotectionScatPctMax}{10.94}
\newcommand{\systemstateprotectionScatPctAvg}{3.32}
\newcommand{\systemstateprotectionScatPctStd}{3.65}
\newcommand{\systemstateprotectionMScatPctMin}{0.03}

\newcommand{\systemstateprotectionMScatPctMedian}{0.65}

\newcommand{\systemstateprotectionMScatPctMax}{7.29}
\newcommand{\systemstateprotectionMScatPctAvg}{1.62}
\newcommand{\systemstateprotectionMScatPctStd}{2.27}

\newcommand{\systemstateprotectionLocSizeQThree}{3.0}
\newcommand{\systemstateprotectionLocSizeMax}{449.0}

\newcommand{\accesscontrolScatAverage}{\accesscontrolScatAvg}
\newcommand{\accesscontrolMScatAverage}{\accesscontrolMScatAvg}
\newcommand{\accesscontrolLLocPctAverage}{\accesscontrolLLocPctAvg}

\newcommand{\cryptographyScatAverage}{\cryptographyScatAvg}
\newcommand{\cryptographyMScatAverage}{\cryptographyMScatAvg}
\newcommand{\cryptographyLLocPctAverage}{\cryptographyLLocPctAvg}

\newcommand{\securitymonitoringScatAverage}{\securitymonitoringScatAvg}
\newcommand{\securitymonitoringMScatAverage}{\securitymonitoringMScatAvg}
\newcommand{\securitymonitoringLLocPctAverage}{\securitymonitoringLLocPctAvg}

\newcommand{\securedatahandlingScatAverage}{\securedatahandlingScatAvg}
\newcommand{\securedatahandlingMScatAverage}{\securedatahandlingMScatAvg}
\newcommand{\securedatahandlingLLocPctAverage}{\securedatahandlingLLocPctAvg}

\newcommand{\systemstateprotectionScatAverage}{\systemstateprotectionScatAvg}
\newcommand{\systemstateprotectionMScatAverage}{\systemstateprotectionMScatAvg}
\newcommand{\systemstateprotectionLLocPctAverage}{\systemstateprotectionLLocPctAvg}

\newcommand{\totalprojects}{9}

\newcommand{\signalserverFiles}{759}
\newcommand{\signalserverTotalLLoc}{48,623}
\newcommand{\signalserverAnnLLocU}{19,368}
\newcommand{\signalserverLLocPct}{49.62}
\newcommand{\signalserverScatAvg}{27.12}
\newcommand{\signalserverTangAvg}{1.65}
\newcommand{\signalserverFeatures}{33}
\newcommand{\dolphinschedulerFiles}{1,963}
\newcommand{\dolphinschedulerTotalLLoc}{101,118}
\newcommand{\dolphinschedulerAnnLLocU}{6,492}
\newcommand{\dolphinschedulerLLocPct}{7.88}
\newcommand{\dolphinschedulerScatAvg}{28.00}
\newcommand{\dolphinschedulerTangAvg}{1.53}
\newcommand{\dolphinschedulerFeatures}{21}
\newcommand{\eurekaFiles}{267}
\newcommand{\eurekaTotalLLoc}{22,347}
\newcommand{\eurekaAnnLLocU}{1,658}
\newcommand{\eurekaLLocPct}{8.18}
\newcommand{\eurekaScatAvg}{5.70}
\newcommand{\eurekaTangAvg}{1.39}
\newcommand{\eurekaFeatures}{23}
\newcommand{\hadoopFiles}{7,832}
\newcommand{\hadoopTotalLLoc}{1,009,462}
\newcommand{\hadoopAnnLLocU}{96,028}
\newcommand{\hadoopLLocPct}{12.63}
\newcommand{\hadoopScatAvg}{123.57}
\newcommand{\hadoopTangAvg}{1.85}
\newcommand{\hadoopFeatures}{35}
\newcommand{\hudiFiles}{2,126}
\newcommand{\hudiTotalLLoc}{209,222}
\newcommand{\hudiAnnLLocU}{622}
\newcommand{\hudiLLocPct}{0.34}
\newcommand{\hudiScatAvg}{6.14}
\newcommand{\hudiTangAvg}{1.63}
\newcommand{\hudiFeatures}{14}
\newcommand{\icebergFiles}{2,171}
\newcommand{\icebergTotalLLoc}{246,733}
\newcommand{\icebergAnnLLocU}{5,565}
\newcommand{\icebergLLocPct}{2.61}
\newcommand{\icebergScatAvg}{17.24}
\newcommand{\icebergTangAvg}{1.50}
\newcommand{\icebergFeatures}{17}
\newcommand{\rocketmqFiles}{1,474}
\newcommand{\rocketmqTotalLLoc}{168,351}
\newcommand{\rocketmqAnnLLocU}{7,906}
\newcommand{\rocketmqLLocPct}{5.62}
\newcommand{\rocketmqScatAvg}{24.23}
\newcommand{\rocketmqTangAvg}{1.47}
\newcommand{\rocketmqFeatures}{13}
\newcommand{\tomcatFiles}{1,548}
\newcommand{\tomcatTotalLLoc}{248,817}
\newcommand{\tomcatAnnLLocU}{42,284}
\newcommand{\tomcatLLocPct}{22.07}
\newcommand{\tomcatScatAvg}{26.29}
\newcommand{\tomcatTangAvg}{1.66}
\newcommand{\tomcatFeatures}{31}
\newcommand{\traccarFiles}{983}
\newcommand{\traccarTotalLLoc}{73,459}
\newcommand{\traccarAnnLLocU}{3,472}
\newcommand{\traccarLLocPct}{5.94}
\newcommand{\traccarScatAvg}{8.88}
\newcommand{\traccarTangAvg}{1.58}
\newcommand{\traccarFeatures}{32}

\newcommand{\sumFiles}{19,123}
\newcommand{\sumTotalLLoc}{2,128,132}
\newcommand{\sumAnnLLocU}{183,395}
\newcommand{\sumLLocPct}{11.15}
\newcommand{\sumFeatures}{561}
\newcommand{\sumDistinctFeatures}{54}
\newcommand{\avgFiles}{2,124.6}
\newcommand{\avgTotalLLoc}{236,417.9}
\newcommand{\avgAnnLLocU}{20,377.2}
\newcommand{\avgScat}{30.56}
\newcommand{\avgTang}{1.57}
\newcommand{\avgFeatures}{22.8}

\newcommand{\rqOneAbsSizeR}{0.92}
\newcommand{\rqOneRelSizeR}{-0.09}
\newcommand{\rqOneAbsScatFilesR}{0.94}
\newcommand{\rqOneAbsScatMethodsR}{0.90}
\newcommand{\rqOneRelScatFilesR}{-0.20}
\newcommand{\rqOneRelScatMethodsR}{-0.30}

\newcommand{\patTotal}{19,712}

\newcommand{\patExtWrapperPct}{3.0}

\newcommand{\patCustImplPct}{22.5}

\newcommand{\patExtUsagePct}{15.9}

\newcommand{\patCustUsagePct}{58.6}

\newcommand{\patUsesTotal}{\patTotal}

\newcommand{\secfnsTotal}{4,351}

\newcommand{\featureDistTotalSecFns}{4,351}

\newcommand{\featureDistAccesscontrolSecFnsPct}{48.0}

\newcommand{\featureDistAccessControlSecFnsPct}{\featureDistAccesscontrolSecFnsPct}

\newcommand{\featureDistCryptographySecFnsPct}{39.3}

\newcommand{\featureDistSecuritymonitoringSecFnsPct}{0.7}

\newcommand{\featureDistSecurityMonitoringSecFnsPct}{\featureDistSecuritymonitoringSecFnsPct}

\newcommand{\featureDistSecuredatahandlingSecFnsPct}{9.0}

\newcommand{\featureDistSecureDataHandlingSecFnsPct}{\featureDistSecuredatahandlingSecFnsPct}

\newcommand{\featureDistSystemstateprotectionSecFnsPct}{3.0}

\newcommand{\featureDistSystemStateProtectionSecFnsPct}{\featureDistSystemstateprotectionSecFnsPct}

\newcommand{\rqThreeOriginExtPct}{30.8}
\newcommand{\rqThreeOriginIntExtPct}{43.1}
\newcommand{\rqThreeOriginPureIntPct}{26.0}

\newcommand{\acOriginIntExtPct}{48.5}
\newcommand{\acOriginPureIntPct}{29.1}

\newcommand{\cryptoOriginExtPct}{44.3}

\newcommand{\monOriginPureIntPct}{39.3}

\newcommand{\sspOriginPureIntPct}{39.9}

\newcommand{\rqThreeExtApiTotalUses}{4,196}
\newcommand{\rqThreeExtApiTotalNamespaces}{73}
\newcommand{\rqThreeExtApiTopTwentyOneUses}{4,084}
\newcommand{\rqThreeExtApiTopTwentyOnePct}{97.3}
\newcommand{\rqThreeJdkSecUses}{2,894}
\newcommand{\rqThreeJdkSecUsesPct}{69.0}
\newcommand{\rqThreeExtApiRootOne}{java.security}
\newcommand{\rqThreeExtApiFnsOne}{91}
\newcommand{\rqThreeExtApiUsesOne}{1,520}
\newcommand{\rqThreeExtApiProjOne}{9}
\newcommand{\rqThreeExtApiRootTwo}{javax.net}
\newcommand{\rqThreeExtApiFnsTwo}{52}
\newcommand{\rqThreeExtApiUsesTwo}{620}
\newcommand{\rqThreeExtApiProjTwo}{9}
\newcommand{\rqThreeExtApiRootThree}{javax.security}
\newcommand{\rqThreeExtApiFnsThree}{62}
\newcommand{\rqThreeExtApiUsesThree}{460}
\newcommand{\rqThreeExtApiProjThree}{4}
\newcommand{\rqThreeExtApiRootFour}{software.amazon.awssdk}
\newcommand{\rqThreeExtApiFnsFour}{173}
\newcommand{\rqThreeExtApiUsesFour}{266}
\newcommand{\rqThreeExtApiProjFour}{5}
\newcommand{\rqThreeExtApiRootFive}{javax.crypto}
\newcommand{\rqThreeExtApiFnsFive}{26}
\newcommand{\rqThreeExtApiUsesFive}{227}
\newcommand{\rqThreeExtApiProjFive}{7}
\newcommand{\rqThreeExtApiRootSix}{org.apache.zookeeper}
\newcommand{\rqThreeExtApiFnsSix}{38}
\newcommand{\rqThreeExtApiUsesSix}{146}
\newcommand{\rqThreeExtApiProjSix}{2}
\newcommand{\rqThreeExtApiRootSeven}{io.dropwizard.auth}
\newcommand{\rqThreeExtApiFnsSeven}{9}
\newcommand{\rqThreeExtApiUsesSeven}{107}
\newcommand{\rqThreeExtApiProjSeven}{1}
\newcommand{\rqThreeExtApiRootEight}{org.apache.commons}
\newcommand{\rqThreeExtApiFnsEight}{101}
\newcommand{\rqThreeExtApiUsesEight}{92}
\newcommand{\rqThreeExtApiProjEight}{6}
\newcommand{\rqThreeExtApiRootNine}{io.netty.handler}
\newcommand{\rqThreeExtApiFnsNine}{41}
\newcommand{\rqThreeExtApiUsesNine}{83}
\newcommand{\rqThreeExtApiProjNine}{5}
\newcommand{\rqThreeExtApiRootTen}{jakarta.ws}
\newcommand{\rqThreeExtApiFnsTen}{55}
\newcommand{\rqThreeExtApiUsesTen}{71}
\newcommand{\rqThreeExtApiProjTen}{3}
\newcommand{\rqThreeExtApiRootEleven}{com.microsoft.azure}
\newcommand{\rqThreeExtApiFnsEleven}{57}
\newcommand{\rqThreeExtApiUsesEleven}{70}
\newcommand{\rqThreeExtApiProjEleven}{1}
\newcommand{\rqThreeExtApiRootTwelve}{java.nio}
\newcommand{\rqThreeExtApiFnsTwelve}{64}
\newcommand{\rqThreeExtApiUsesTwelve}{67}
\newcommand{\rqThreeExtApiProjTwelve}{8}
\newcommand{\rqThreeExtApiRootThirteen}{io.grpc}
\newcommand{\rqThreeExtApiFnsThirteen}{41}
\newcommand{\rqThreeExtApiUsesThirteen}{58}
\newcommand{\rqThreeExtApiProjThirteen}{2}
\newcommand{\rqThreeExtApiRootFourteen}{java.sql}
\newcommand{\rqThreeExtApiFnsFourteen}{10}
\newcommand{\rqThreeExtApiUsesFourteen}{57}
\newcommand{\rqThreeExtApiProjFourteen}{5}
\newcommand{\rqThreeExtApiRootFifteen}{javax.servlet}
\newcommand{\rqThreeExtApiFnsFifteen}{26}
\newcommand{\rqThreeExtApiUsesFifteen}{50}
\newcommand{\rqThreeExtApiProjFifteen}{2}
\newcommand{\rqThreeExtApiRootSixteen}{com.amazonaws.auth}
\newcommand{\rqThreeExtApiFnsSixteen}{13}
\newcommand{\rqThreeExtApiUsesSixteen}{37}
\newcommand{\rqThreeExtApiProjSixteen}{3}
\newcommand{\rqThreeExtApiRootSeventeen}{java.net}
\newcommand{\rqThreeExtApiFnsSeventeen}{59}
\newcommand{\rqThreeExtApiUsesSeventeen}{33}
\newcommand{\rqThreeExtApiProjSeventeen}{8}
\newcommand{\rqThreeExtApiRootEighteen}{com.qcloud.cos}
\newcommand{\rqThreeExtApiFnsEighteen}{9}
\newcommand{\rqThreeExtApiUsesEighteen}{33}
\newcommand{\rqThreeExtApiProjEighteen}{2}
\newcommand{\rqThreeExtApiRootNineteen}{org.immutables.value}
\newcommand{\rqThreeExtApiFnsNineteen}{6}
\newcommand{\rqThreeExtApiUsesNineteen}{30}
\newcommand{\rqThreeExtApiProjNineteen}{1}
\newcommand{\rqThreeExtApiRootTwenty}{javax.naming}
\newcommand{\rqThreeExtApiFnsTwenty}{41}
\newcommand{\rqThreeExtApiUsesTwenty}{29}
\newcommand{\rqThreeExtApiProjTwenty}{4}
\newcommand{\rqThreeExtApiRootTwentyOne}{org.apache.flink}
\newcommand{\rqThreeExtApiFnsTwentyOne}{11}
\newcommand{\rqThreeExtApiUsesTwentyOne}{28}
\newcommand{\rqThreeExtApiProjTwentyOne}{1}

\newcommand{\acPatExtWrapperPct}{3.1}

\newcommand{\acPatCustImplPct}{22.0}

\newcommand{\acPatExtUsagePct}{18.3}

\newcommand{\acPatCustUsagePct}{56.6}

\newcommand{\cryptoPatExtWrapperPct}{3.9}

\newcommand{\cryptoPatCustImplPct}{29.5}

\newcommand{\cryptoPatExtUsagePct}{12.0}

\newcommand{\cryptoPatCustUsagePct}{54.6}

\newcommand{\monPatCustImplPct}{3.9}

\newcommand{\monPatCustUsagePct}{88.3}

\newcommand{\sdhPatExtUsagePct}{14.3}

\newcommand{\sspPatCustImplPct}{26.5}

\newcommand{\sspPatExtUsagePct}{12.6}

\newcommand{\sspPatCustUsagePct}{59.8}

\newcommand{\acPatCustPct}{\acPatCustUsagePct}

\newcommand{\monPatLibPct}{7.8}

\newcommand{\monPatCustPct}{\monPatCustUsagePct}

\newcommand{\accesscontrolCustImplSizeMax}{1,437}

\newcommand{\cryptographyCustImplSizeMedian}{6}

\newcommand{\cryptographyCustImplSizeMax}{929}

\newcommand{\securitymonitoringCustImplSizeMax}{1,218}

\newcommand{\securedatahandlingCustImplSizeMedian}{9}

\newcommand{\securedatahandlingCustImplSizeMax}{721}

\newcommand{\systemstateprotectionCustImplSizeMax}{449}

%% file: generated_tangling_macros.tex
\newcommand{\tanglingACtoAC}{11.40}
\newcommand{\tanglingACtoCR}{8.49}
\newcommand{\tanglingACtoSM}{3.10}
\newcommand{\tanglingACtoSD}{4.72}
\newcommand{\tanglingACtoSS}{0.64}
\newcommand{\tanglingACtoSF}{27.20}
\newcommand{\tanglingCRtoAC}{9.28}
\newcommand{\tanglingCRtoCR}{19.21}
\newcommand{\tanglingCRtoSM}{2.54}
\newcommand{\tanglingCRtoSD}{2.52}
\newcommand{\tanglingCRtoSS}{1.02}
\newcommand{\tanglingCRtoSF}{23.80}
\newcommand{\tanglingSMtoAC}{12.13}
\newcommand{\tanglingSMtoCR}{3.50}
\newcommand{\tanglingSMtoSM}{0.00}
\newcommand{\tanglingSMtoSD}{0.23}
\newcommand{\tanglingSMtoSS}{1.04}
\newcommand{\tanglingSMtoSF}{16.44}
\newcommand{\tanglingSDtoAC}{35.25}
\newcommand{\tanglingSDtoCR}{19.38}
\newcommand{\tanglingSDtoSM}{2.85}
\newcommand{\tanglingSDtoSD}{1.62}
\newcommand{\tanglingSDtoSS}{0.82}
\newcommand{\tanglingSDtoSF}{50.39}
\newcommand{\tanglingSStoAC}{4.51}
\newcommand{\tanglingSStoCR}{5.55}
\newcommand{\tanglingSStoSM}{1.51}
\newcommand{\tanglingSStoSD}{0.26}
\newcommand{\tanglingSStoSS}{1.61}
\newcommand{\tanglingSStoSF}{14.62}
\newcommand{\tanglingSFtoAC}{12.71}
\newcommand{\tanglingSFtoCR}{8.69}
\newcommand{\tanglingSFtoSM}{2.39}
\newcommand{\tanglingSFtoSD}{3.48}
\newcommand{\tanglingSFtoSS}{0.86}
\newcommand{\tanglingSFtoSF}{21.12}

\definecolor{cellACAC}{RGB}{178,207,181}
\definecolor{cellACCR}{RGB}{191,216,194}
\definecolor{cellACSM}{RGB}{216,234,219}
\definecolor{cellACSD}{RGB}{209,229,211}
\definecolor{cellACSS}{RGB}{228,242,230}
\definecolor{cellACSF}{RGB}{104,154,110}
\definecolor{cellCRAC}{RGB}{187,214,191}
\definecolor{cellCRCR}{RGB}{141,181,146}
\definecolor{cellCRSM}{RGB}{219,236,221}
\definecolor{cellCRSD}{RGB}{219,236,221}
\definecolor{cellCRSS}{RGB}{226,241,228}
\definecolor{cellCRSF}{RGB}{120,166,125}
\definecolor{cellSMAC}{RGB}{174,204,178}
\definecolor{cellSMCR}{RGB}{214,233,217}
\definecolor{cellSMSM}{RGB}{231,245,233}
\definecolor{cellSMSD}{RGB}{229,244,231}
\definecolor{cellSMSS}{RGB}{226,241,228}
\definecolor{cellSMSF}{RGB}{154,190,158}
\definecolor{cellSDAC}{RGB}{67,128,73}
\definecolor{cellSDCR}{RGB}{140,180,145}
\definecolor{cellSDSM}{RGB}{217,235,220}
\definecolor{cellSDSD}{RGB}{223,239,225}
\definecolor{cellSDSS}{RGB}{227,242,229}
\definecolor{cellSDSF}{RGB}{22,96,30}
\definecolor{cellSSAC}{RGB}{210,230,212}
\definecolor{cellSSCR}{RGB}{205,226,207}
\definecolor{cellSSSM}{RGB}{223,240,226}
\definecolor{cellSSSD}{RGB}{229,244,231}
\definecolor{cellSSSS}{RGB}{223,239,225}
\definecolor{cellSSSF}{RGB}{163,196,167}
\definecolor{cellSFAC}{RGB}{171,202,175}
\definecolor{cellSFCR}{RGB}{190,216,193}
\definecolor{cellSFSM}{RGB}{219,237,222}
\definecolor{cellSFSD}{RGB}{214,233,217}
\definecolor{cellSFSS}{RGB}{227,242,229}
\definecolor{cellSFSF}{RGB}{132,175,137}

\definecolor{cellOccACAC}{RGB}{201,224,204}
\definecolor{cellOccACCR}{RGB}{208,228,210}
\definecolor{cellOccACSM}{RGB}{214,233,216}
\definecolor{cellOccACSD}{RGB}{217,235,220}
\definecolor{cellOccACSS}{RGB}{226,242,229}
\definecolor{cellOccACSF}{RGB}{154,190,158}
\definecolor{cellOccCRAC}{RGB}{171,202,175}
\definecolor{cellOccCRCR}{RGB}{216,234,219}
\definecolor{cellOccCRSM}{RGB}{208,229,211}
\definecolor{cellOccCRSD}{RGB}{212,232,215}
\definecolor{cellOccCRSS}{RGB}{222,239,225}
\definecolor{cellOccCRSF}{RGB}{132,174,137}
\definecolor{cellOccSMAC}{RGB}{172,203,175}
\definecolor{cellOccSMCR}{RGB}{221,238,223}
\definecolor{cellOccSMSM}{RGB}{231,245,233}
\definecolor{cellOccSMSD}{RGB}{228,243,230}
\definecolor{cellOccSMSS}{RGB}{229,244,231}
\definecolor{cellOccSMSF}{RGB}{161,195,165}
\definecolor{cellOccSDAC}{RGB}{73,132,80}
\definecolor{cellOccSDCR}{RGB}{169,201,173}
\definecolor{cellOccSDSM}{RGB}{214,233,216}
\definecolor{cellOccSDSD}{RGB}{229,243,231}
\definecolor{cellOccSDSS}{RGB}{222,238,224}
\definecolor{cellOccSDSF}{RGB}{40,109,48}
\definecolor{cellOccSSAC}{RGB}{183,211,187}
\definecolor{cellOccSSCR}{RGB}{223,239,225}
\definecolor{cellOccSSSM}{RGB}{218,236,220}
\definecolor{cellOccSSSD}{RGB}{226,241,228}
\definecolor{cellOccSSSS}{RGB}{231,245,233}
\definecolor{cellOccSSSF}{RGB}{170,201,173}
\definecolor{cellOccSFAC}{RGB}{164,197,168}
\definecolor{cellOccSFCR}{RGB}{186,213,189}
\definecolor{cellOccSFSM}{RGB}{192,217,195}
\definecolor{cellOccSFSD}{RGB}{215,233,217}
\definecolor{cellOccSFSS}{RGB}{222,239,224}
\definecolor{cellOccSFSF}{RGB}{51,116,58}

\newcommand{\tanglingSymACtoAC}{11.40}
\newcommand{\tanglingSymACtoCR}{4.90}
\newcommand{\tanglingSymACtoSM}{0.88}
\newcommand{\tanglingSymACtoSD}{4.92}
\newcommand{\tanglingSymACtoSS}{0.34}
\newcommand{\tanglingSymACtoSF}{11.03}
\newcommand{\tanglingSymCRtoAC}{16.09}
\newcommand{\tanglingSymCRtoCR}{19.21}
\newcommand{\tanglingSymCRtoSM}{0.48}
\newcommand{\tanglingSymCRtoSD}{5.13}
\newcommand{\tanglingSymCRtoSS}{0.79}
\newcommand{\tanglingSymCRtoSF}{22.48}
\newcommand{\tanglingSymSMtoAC}{42.95}
\newcommand{\tanglingSymSMtoCR}{18.57}
\newcommand{\tanglingSymSMtoSM}{0.00}
\newcommand{\tanglingSymSMtoSD}{5.51}
\newcommand{\tanglingSymSMtoSS}{1.57}
\newcommand{\tanglingSymSMtoSF}{68.60}
\newcommand{\tanglingSymSDtoAC}{33.81}
\newcommand{\tanglingSymSDtoCR}{9.53}
\newcommand{\tanglingSymSDtoSM}{0.12}
\newcommand{\tanglingSymSDtoSD}{1.62}
\newcommand{\tanglingSymSDtoSS}{0.14}
\newcommand{\tanglingSymSDtoSF}{43.60}
\newcommand{\tanglingSymSStoAC}{8.58}
\newcommand{\tanglingSymSStoCR}{7.23}
\newcommand{\tanglingSymSStoSM}{1.01}
\newcommand{\tanglingSymSStoSD}{1.54}
\newcommand{\tanglingSymSStoSS}{1.61}
\newcommand{\tanglingSymSStoSF}{18.36}
\newcommand{\tanglingSymSFtoAC}{14.99}
\newcommand{\tanglingSymSFtoCR}{6.92}
\newcommand{\tanglingSymSFtoSM}{0.65}
\newcommand{\tanglingSymSFtoSD}{3.88}
\newcommand{\tanglingSymSFtoSS}{0.60}
\newcommand{\tanglingSymSFtoSF}{21.12}

\definecolor{cellSymACAC}{RGB}{196,220,199}
\definecolor{cellSymACCR}{RGB}{216,234,218}
\definecolor{cellSymACSM}{RGB}{228,243,230}
\definecolor{cellSymACSD}{RGB}{216,234,218}
\definecolor{cellSymACSS}{RGB}{229,244,232}
\definecolor{cellSymACSF}{RGB}{198,221,201}
\definecolor{cellSymCRAC}{RGB}{182,210,186}
\definecolor{cellSymCRCR}{RGB}{173,204,177}
\definecolor{cellSymCRSM}{RGB}{229,243,231}
\definecolor{cellSymCRSD}{RGB}{215,234,218}
\definecolor{cellSymCRSS}{RGB}{228,243,230}
\definecolor{cellSymCRSF}{RGB}{163,197,167}
\definecolor{cellSymSMAC}{RGB}{102,153,108}
\definecolor{cellSymSMCR}{RGB}{175,205,179}
\definecolor{cellSymSMSM}{RGB}{231,245,233}
\definecolor{cellSymSMSD}{RGB}{214,233,217}
\definecolor{cellSymSMSS}{RGB}{226,241,228}
\definecolor{cellSymSMSF}{RGB}{26,98,34}
\definecolor{cellSymSDAC}{RGB}{130,173,134}
\definecolor{cellSymSDCR}{RGB}{202,224,205}
\definecolor{cellSymSDSM}{RGB}{230,244,232}
\definecolor{cellSymSDSD}{RGB}{226,241,228}
\definecolor{cellSymSDSS}{RGB}{230,244,232}
\definecolor{cellSymSDSF}{RGB}{100,152,106}
\definecolor{cellSymSSAC}{RGB}{205,226,208}
\definecolor{cellSymSSCR}{RGB}{209,229,212}
\definecolor{cellSymSSSM}{RGB}{227,242,230}
\definecolor{cellSymSSSD}{RGB}{226,241,228}
\definecolor{cellSymSSSS}{RGB}{226,241,228}
\definecolor{cellSymSSSF}{RGB}{176,205,179}
\definecolor{cellSymSFAC}{RGB}{186,213,189}
\definecolor{cellSymSFCR}{RGB}{210,230,212}
\definecolor{cellSymSFSM}{RGB}{229,243,231}
\definecolor{cellSymSFSD}{RGB}{219,236,221}
\definecolor{cellSymSFSS}{RGB}{229,243,231}
\definecolor{cellSymSFSF}{RGB}{167,200,171}

\definecolor{cellOccSymACAC}{RGB}{220,237,223}
\definecolor{cellOccSymACCR}{RGB}{217,235,219}
\definecolor{cellOccSymACSM}{RGB}{223,239,225}
\definecolor{cellOccSymACSD}{RGB}{222,238,224}
\definecolor{cellOccSymACSS}{RGB}{228,243,230}
\definecolor{cellOccSymACSF}{RGB}{185,212,188}
\definecolor{cellOccSymCRAC}{RGB}{181,209,185}
\definecolor{cellOccSymCRCR}{RGB}{226,241,228}
\definecolor{cellOccSymCRSM}{RGB}{221,238,224}
\definecolor{cellOccSymCRSD}{RGB}{219,236,221}
\definecolor{cellOccSymCRSS}{RGB}{227,242,229}
\definecolor{cellOccSymCRSF}{RGB}{130,173,134}
\definecolor{cellOccSymSMAC}{RGB}{153,189,157}
\definecolor{cellOccSymSMCR}{RGB}{206,227,209}
\definecolor{cellOccSymSMSM}{RGB}{231,245,233}
\definecolor{cellOccSymSMSD}{RGB}{226,241,228}
\definecolor{cellOccSymSMSS}{RGB}{228,242,230}
\definecolor{cellOccSymSMSF}{RGB}{22,96,30}
\definecolor{cellOccSymSDAC}{RGB}{114,161,119}
\definecolor{cellOccSymSDCR}{RGB}{185,212,188}
\definecolor{cellOccSymSDSM}{RGB}{224,240,226}
\definecolor{cellOccSymSDSD}{RGB}{230,244,232}
\definecolor{cellOccSymSDSS}{RGB}{226,242,229}
\definecolor{cellOccSymSDSF}{RGB}{88,143,94}
\definecolor{cellOccSymSSAC}{RGB}{189,215,192}
\definecolor{cellOccSymSSCR}{RGB}{214,233,216}
\definecolor{cellOccSymSSSM}{RGB}{225,241,227}
\definecolor{cellOccSymSSSD}{RGB}{225,241,227}
\definecolor{cellOccSymSSSS}{RGB}{231,245,233}
\definecolor{cellOccSymSSSF}{RGB}{144,183,148}
\definecolor{cellOccSymSFAC}{RGB}{202,224,205}
\definecolor{cellOccSymSFCR}{RGB}{216,234,218}
\definecolor{cellOccSymSFSM}{RGB}{223,239,225}
\definecolor{cellOccSymSFSD}{RGB}{222,238,224}
\definecolor{cellOccSymSFSS}{RGB}{228,243,230}
\definecolor{cellOccSymSFSF}{RGB}{157,192,161}

%% file: 01_Introduction.tex
\section{Introduction}
Modern software systems implement a wide range of security-specific functionality, commonly referred to as security features \cite{hermann2025taxonomy}.
Applications authenticate users and services, enforce authorization decisions, protect data at rest and in transit, manage credentials and cryptographic keys, validate system state, and record security-relevant events. 
Yet, implementing such functionality correctly remains difficult. 
Empirical studies have repeatedly shown that developers struggle to implement security-sensitive functionality correctly, including cryptographic API usage and password storage \cite{nadi2016jumping,acar2017crypto,naiakshina2017password}.
Even when developers actively seek implementation guidance, the resources they consult can substantially affect the security of the resulting code \cite{acar2016information}. 
At scale, incorrect integration of otherwise well-established security mechanisms can lead to exploitable vulnerabilities, as demonstrated, for example, for TLS in Android applications \cite{fahl2012eve}.

One response to these difficulties is to encapsulate security functionality in libraries and frameworks. 
Such abstractions provide reusable implementations of cryptographic primitives, authentication mechanisms, access-control infrastructure, secure communication, and other security functionality, reducing the need for application developers to implement sensitive mechanisms from scratch. 
However, using a security library does not eliminate the application-specific code required to integrate a security mechanism. 
Studies of security APIs show that developers often require substantial knowledge to compose and configure them correctly \cite{nadi2016jumping,acar2017crypto,krueger2021python}.
Security libraries and frameworks therefore provide useful entry points for locating security functionality, but they do not necessarily capture all code involved in realizing a security feature \cite{hermann2025taxonomy}.

Prior empirical work has also examined the usability and misuse of security APIs \cite{nadi2016jumping,acar2017crypto,krueger2021python}, showing that developers frequently struggle with security-sensitive APIs and their correct integration. 
At the same time, feature-location\,\cite{rubin2013survey,bernardi2016crosscutting} research has shown that scattering and tangling complicate program comprehension and maintenance.
For security features, however, it remains unclear how much code they comprise, how widely they are scattered, how strongly they are tangled with other security features, and what part of them comprises custom code instead of library uses.

Despite extensive research on vulnerabilities, security APIs, secure development practices, and security-relevant code \cite{hermann2026toolsexploitablevulnerabilitiessurvey,tahaei2019survey,green2016developers,van2020schrodinger,piskachev2019codebase,Johnson2024,krishnamurthy2026detection}, we lack a systematic understanding of how functional security features are realized across complete software systems.
In particular, it remains unclear how much code they comprise, how widely they are distributed throughout a codebase, which parts are implemented within a project rather than provided by external functionality, and how different
security features interact.
Understanding these characteristics is important for techniques that aim to locate, analyze, maintain, or reason about security functionality.

To address these open questions, we investigate four research questions:

\begin{itemize}
	\item [\textbf{RQ1:}] What is the code-level footprint of security features?
	\item [\textbf{RQ2:}] What are implementation and usage patterns of security features? 
	\item [\textbf{RQ3}:] What parts of security features are implemented within a software project?
	\item [\textbf{RQ4:}] How do security features interact?
\end{itemize}

We conduct an empirical study of functional security features in nine mature Java systems from seven categories drawn from GitHub, comprising 2,127,761 lines of code across 19,121 files. 
We manually identify and annotate security-feature code at both line and block granularity and map the identified functionality to a taxonomy of functional security features derived by Hermann et al.\ \cite{hermann2025taxonomy}.
The resulting dataset contains \sumFeatures\ security features across the \sumAnnLLocU\ annotated lines of code in \patUsesTotal\ locations and records their source-code footprint, including size, distribution, interactions with other security features, and use of internal and external functionality. We contribute:

\begin{itemize}
	\item a curated dataset of \sumFeatures\ functional security features from \totalprojects\ mature Java software systems;
	\item a characterization of the size, scattering, and tangling of security features; and
	\item an analysis of custom implemented and externally provided security functionality.
\end{itemize}

Our analysis shows that security features are more than just small wrappers around external security libraries.
Across the studied systems, most security-relevant functionality is implemented internally, although custom implementations frequently use external functionality.
The balance differs considerably among security feature categories. 
Standardized functionality, such as cryptographic primitives, is more likely to be externalized. 
In contrast, application-specific functionality, such as session management and access-control logic, requires substantially more project-specific integration and implementation. 
Security features also differ considerably in their footprint and distribution throughout the codebase. 
Access-control functionality, in particular, spans large portions of some systems. 
Analyses that identify security functionality solely through recognizable security-library calls therefore miss parts of its implementation.

%% file: 02_Background.tex
\section{Background}
We give an overview of security features, their characteristics, and our annotation approach.

\parhead{Functional Security Features.}
Security features are functionalities of a software system that enforce security properties, such as protecting the integrity of a system against malicious actions or preserving the confidentiality of sensitive data\,\cite{Tsipenyuk2005,Allen2008,mcgraw2004software,Potter2004,hermann2025engineering,hermann2025taxonomy}.
Examples include authentication, authorization, encryption, secure storage, and security monitoring.
Such functionality must be selected, implemented, and maintained as part of the broader software system\,\cite{hermann2025engineering,mcgraw2004software,Potter2004}.
This can be challenging because developers are often not security experts and must correctly integrate security mechanisms provided by libraries and frameworks into application-specific code\,\cite{naiakshina2017password,acar2017crypto,acar2016information,nadi2016jumping}.

To systematically characterize security functionality, we build on the taxonomy of Hermann et al.\,\cite{hermann2025taxonomy}.
The taxonomy comprises 68 implementation-level functional security features organized into five top-level categories and provides guidance for locating their realizations in code:

\begin{description}
\item [Access Control.]
Authentication and authorization functionality for identifying users or services and controlling access to system resources.

\item [Cryptography.]
Functionality for protecting data using encryption, hashing, key management, and related cryptographic mechanisms.

\item [Security Monitoring.]
Functionality for recording and observing security-relevant events, including logging, auditing, and intrusion detection.

\item [Secure Data Handling.]
Functionality for securely storing and processing sensitive or untrusted data, including input validation and protection against injection attacks.

\item [System State Protection.]
Functionality for preserving the integrity and availability of system state, including session management, synchronization, and resource management.
\end{description}

Hermann et al.\,\cite{hermann2025taxonomy} show that these features can be exposed through framework APIs, annotations, and configuration mechanisms.
We use their taxonomy as the conceptual basis for identifying and categorizing security-feature implementations in complete software systems.

\parhead{Feature Annotations.}
To record feature locations in source code, we use the embedded feature annotation notation\,\cite{Schwarz2020}.
The notation marks either contiguous blocks of code or individual lines with a feature name, which is defined in a feature model.
A code block is delimited by \texttt{//\&begin[FeatureName]} and \texttt{//\&end[FeatureName]}, while individual lines are annotated using \texttt{//\&line[FeatureName]}.
The notation represents feature locations directly in the source code and allows annotations of multiple features to overlap.
For example, the code excerpt at the bottom of \cref{fig:methodology} shows the use of these annotations to annotate lines 236--280 as belonging to the security feature \texttt{Password}, and the line 232 as also belonging to the security feature \texttt{Salting}.

\parhead{Feature Scattering and Tangling.}
Software features are not necessarily localized in a single module or component.
A feature is \emph{scattered} when its implementation is distributed across multiple program elements, while code is \emph{tangled} when the same program element contributes to multiple features or concerns\,\cite{bernardi2016crosscutting,Eaddy2008}.
Scattering is commonly measured using concern-diffusion metrics, such as Concern Diffusion over Components (CDC) and Concern Diffusion over Operations (CDO), which count the number of components or operations contributing to a concern\,\cite{SantAnna2003,Eaddy2008}.
In our study, we analogously measure scattering by the number and relative share of files and methods containing a security feature.
We measure tangling through overlapping feature annotations, quantifying how much code associated with one security feature is also associated with another.

Scattering and tangling can complicate program comprehension and maintenance, as developers need to identify and reason about multiple code locations to understand or modify a feature.
For security features, these properties are particularly relevant because security functionality can be distributed throughout the system and overlap with other security functionality.
We therefore characterize how widely security features are distributed across the codebase and their overlap.

%% file: 03_Methodology.tex
\section{Methodology}

\begin{figure*}
    \includegraphics[width=.85\textwidth]{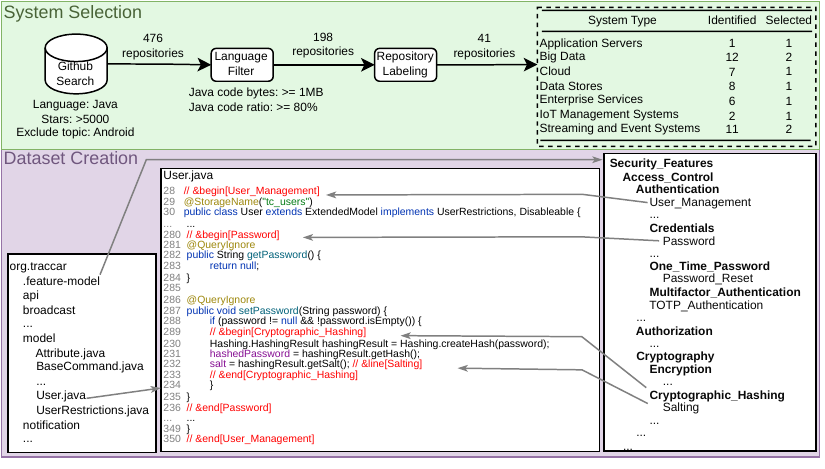}
	\vspace{-10pt}
    \caption{Overview of our methodology}
    \label{fig:methodology}
	\vspace{-10pt}
\end{figure*}

We conducted a repository mining study of security features in Java software systems.
We oriented our study on the ACM empirical standards for conducting and reporting repository mining studies\,\cite{acm_empirical_standards_repository_mining}.
Our methodology consists of three main steps: selecting suitable repositories, identifying and annotating security feature code, and analyzing the resulting dataset with respect to our research questions.
\Cref{fig:methodology} provides an overview of our methodology.

\subsection{Repository Filtering}

Our study requires repositories that represent complete, mature software systems in which security functionality is embedded in a broader application context.
We are interested in how security features are implemented as part of a system, how widely their code is distributed, how they interact with each other, and to what extent they rely on custom implementation.


We are interested in complete general-purpose software systems that integrate security features as part of their overall functionality, rather than libraries that provide them.
To identify suitable subject systems for our analysis, we therefore apply the following selection criteria:
\begin{enumerate}
	\item \textbf{System Type:}
	The system is a complete software system. 
	Complete systems provide the application context in which security features interact with other system functionality.

	\item \textbf{Security Focus:}
	Security is not the primary purpose of the system.
	This avoids selecting systems in which security functionality is inherently dominant and, therefore, not representative of how security features occur in general-purpose software systems.

	\item \textbf{Maintenance Status:}
	The repository is actively maintained.
	This limits the study to systems that are still developed and avoids analyzing obsolete or abandoned implementations.

	\item \textbf{Repository Independence:}
	The repository is not a fork of another project.
	This avoids treating substantially duplicated codebases as independent subject systems.

	\item \textbf{Documentation Language:}
	The repository's README is written in English.
	We rely on project documentation during repository selection to determine its purpose.

	\item \textbf{Implementation Language:}
	Java is the dominant programming language of the repository.
	Restricting to one language improves comparability between systems and avoids language-specific implementation differences in our analysis.
	Java also provides a sufficiently large set of mature server-side and infrastructure systems from different application domains.

	\item \textbf{Codebase Size:}
	The repository contains a sufficiently large Java codebase to study the distribution and implementation structure of security features.
	Small Java components embedded in otherwise non-Java systems are therefore excluded.

	\item \textbf{Repository Popularity:}
	The repository has substantial community adoption.
	We use popularity as an initial proxy for selecting established projects to avoid rarely used repositories.
\end{enumerate}

\noindent
We operationalized these criteria in three filtering steps.
First, we queried the GitHub Search API for repositories containing Java code (6) that have at least 5,000 stars (8).
We excluded repositories labeled as ``Android,'' as these follow a platform-specific security architecture\,\cite{Mayrhofer2021} that differs substantially from general-purpose systems.
This resulted in 467 repositories.
Second, we queried the GitHub Languages API and retained those containing at least 1 MB of Java code (7) for which Java accounts for at least 80\% of the repository's source code (6).
This reduced the set to 198 repositories.
Finally, the first author manually inspected each remaining repository and applied the application-oriented (1), security-focus (2), maintenance (3), fork (4), and documentation language (5) criteria.
After this step, 41 repositories remained as candidates for analysis.

\subsection{Repository Selection}

Our analysis requires detailed examination of security features within each subject system.
We therefore selected a subset of repositories to enable in-depth analysis while preserving the diversity of system types in the candidate pool.
To characterize this diversity, we examined the purpose and functionality of each repository.
We iteratively grouped repositories according to these characteristics.
After four rounds of discussion, we identified 7 system categories.
Repositories could be assigned to multiple categories when their functionality spanned more than one system type:

\smallskip
\begin{minipage}{0.4\linewidth}
	\begin{itemize}
	    \item Application Servers (1 system)
	    \item Big Data (12 systems)
	    \item Cloud (7 systems)
	    \item Data Stores (8 systems)
	\end{itemize}
\end{minipage}
\begin{minipage}{0.6\linewidth}
	\begin{itemize}
	    \item Enterprise Services (6 systems)
		\item IoT Management Systems (2 systems)
		\item Streaming and Event Systems (11 systems)
		\item[\vspace{\fill}]
	\end{itemize}
\end{minipage}
\smallskip

\noindent
We selected 9 repositories for detailed analysis, covering at least one repository of each system type.
For system types with more than 10 systems, we included two repositories to better represent the diversity of our candidate pool.
We further selected systems of different sizes to capture variation between smaller and larger codebases.
\Cref{tab:project-overview} shows the final set of selected repositories.

\subsection{Security Feature Identification and Annotation}

We systematically identified and annotated security features in the selected repositories.

\subsubsection{Scope}
Since we are interested in security features written and used by developers, we analyzed all Java production files of the repositories.
Therefore, we excluded tests, benchmarks, other embedded applications, dependencies, examples, documentation, resources, and websites contained in the repository, as they do not provide logic for the application's operation.
However, we used these artifacts to gain a better understanding of the system and its security features.

We recorded security features by annotating singular lines or blocks of code if they contributed to the implementation or usage of a security feature.
This could include, e.g., a method, a class, or a set of lines of code within a method that related to a security feature.
The annotation process used a standardized notation\,\cite{Schwarz2020} and HAnS as a corresponding annotation editor\,\cite{Martinson2021}.

\subsubsection{Annotation Process}
The first author, a PhD student with four years of experience in software security, manually investigated the codebase of the repositories and annotated all lines of code that implement or use a security feature.
We used the taxonomy of Hermann et al.\,\cite{hermann2025taxonomy} for our analysis, since they provide a structured collection of security features and guidance on how they can be located.
We mapped all identified security features to this taxonomy.
To better understand how security features are realized in practice, we recorded identified features at a finer level of granularity than provided by the taxonomy.
For example, we recorded authentication tokens or external service credentials, which are mapped to the security feature credentials in the taxonomy.

The first author was highly inclusive in the annotation of all repositories to improve recall, i.e., annotating all potential security feature code, even when not certain about its classification.
The first and second authors subsequently reviewed the dataset by examining the annotated code, and discussed the taxonomy mapping.
They resolved any disagreements through discussion until a consensus was reached.
We chose the timing and frequency of reviews based on the repository size and the number of its modules.
However, in the early stages of the annotation process, we held more frequent discussions to ensure a shared understanding.
This process was repeated until all repositories were fully annotated to ensure the validity and the reliability of our dataset.
The whole process took approximately 6 months to complete.

As a final validation step, we used our findings of our study to identify potential gaps or inconsistencies in our annotations and the mapping to the taxonomy.
For example, we used internal and external APIs we categorized as contributing to security features (RQ3) to cross-check whether the annotations in the code were consistent and complete.
The precision of our annotations around API usage was already high (92.8\%), with only a few minor inconsistencies that we corrected.

\subsubsection{Taxonomy Refinement}
In total, we identified \sumFeatures\ security features across all analyzed repositories.
While we already sorted them according to the existing taxonomy, the first and second author refined the initial categorization in two separate meetings to ensure that our taxonomy accurately reflects the security features as we observed them in the projects.
While we were able to categorize all security features according to the original taxonomy, some security features required discussions on their proper placement within the taxonomy, as they could fit into multiple categories.
We carefully reviewed and discussed these cases until a consensus was reached.
This resulted in 12 features being moved to different categories, and 14 features moved within the same category.

\subsection{Data Analysis}

We analyzed the annotated security features at the level of the predefined taxonomy with respect to the four research questions.
For RQ1, we characterized the implementation structure of security features by measuring their size, and scattering across the codebase.
For RQ2, we classified how security features are implemented, distinguishing custom implementation from functionality obtained through reusable security libraries.
For RQ3, we analyzed whether security-related functionality is provided internally by the project or through external APIs.
For RQ4, we examined how security features interact with each other by analyzing their shared lines of code.

%% file: 04_Dataset.tex
\subsection{Dataset}

Recall, our dataset comprises 9 open-source systems from 7 different categories.
We analyzed \sumTotalLLoc\ lines of code in \sumFiles\ files, of which \sumAnnLLocU\ lines implement \sumFeatures\ security features related to \sumDistinctFeatures\ security features from the taxonomy.
We use the taxonomy features for our analysis.

\begin{table*}[t]
\centering
\caption{Overview of subject systems}
\vspace{-10pt}
\label{tab:project-overview}
\smaller
\setlength{\tabcolsep}{4pt}

\begin{tabular}{lrrrrrrrr}
\toprule
\textbf{Project} & \textbf{Version} & \textbf{Files} & \textbf{Total LLOC} & \textbf{Ann LLOC} & \textbf{LLOC\%} & \textbf{Avg Scat} & \textbf{Avg Tang} & \textbf{Features} \\
\midrule
Signal-Server & v20250627.0.0 & \signalserverFiles & \signalserverTotalLLoc & \signalserverAnnLLocU & \signalserverLLocPct\% & \signalserverScatAvg & \signalserverTangAvg & \signalserverFeatures \\
dolphinscheduler & 3.3.0 & \dolphinschedulerFiles & \dolphinschedulerTotalLLoc & \dolphinschedulerAnnLLocU & \dolphinschedulerLLocPct\% & \dolphinschedulerScatAvg & \dolphinschedulerTangAvg & \dolphinschedulerFeatures \\
eureka & 2.0.5 & \eurekaFiles & \eurekaTotalLLoc & \eurekaAnnLLocU & \eurekaLLocPct\% & \eurekaScatAvg & \eurekaTangAvg & \eurekaFeatures \\
hadoop & 3.4.2-RC1 & \hadoopFiles & \hadoopTotalLLoc & \hadoopAnnLLocU & \hadoopLLocPct\% & \hadoopScatAvg & \hadoopTangAvg & \hadoopFeatures \\
hudi & 1.0.2 & \hudiFiles & \hudiTotalLLoc & \hudiAnnLLocU & \hudiLLocPct\% & \hudiScatAvg & \hudiTangAvg & \hudiFeatures \\
iceberg & 1.9.x & \icebergFiles & \icebergTotalLLoc & \icebergAnnLLocU & \icebergLLocPct\% & \icebergScatAvg & \icebergTangAvg & \icebergFeatures \\
rocketmq & 5.3.3 & \rocketmqFiles & \rocketmqTotalLLoc & \rocketmqAnnLLocU & \rocketmqLLocPct\% & \rocketmqScatAvg & \rocketmqTangAvg & \rocketmqFeatures \\
tomcat & 11.0.8 & \tomcatFiles & \tomcatTotalLLoc & \tomcatAnnLLocU & \tomcatLLocPct\% & \tomcatScatAvg & \tomcatTangAvg & \tomcatFeatures \\
traccar & 6.6 & \traccarFiles & \traccarTotalLLoc & \traccarAnnLLocU & \traccarLLocPct\% & \traccarScatAvg & \traccarTangAvg & \traccarFeatures \\
\midrule
AVERAGE &  & \avgFiles & \avgTotalLLoc & \avgAnnLLocU & \sumLLocPct\% & \avgScat & \avgTang & \avgFeatures \\
\bottomrule
\end{tabular}
\vspace{-14pt}
\end{table*}

\parhead{Apache DolphinScheduler (Streaming and Event System)} is a workflow scheduling platform for complex data pipelines\,\cite{dolphinscheduler}.
It implements \dolphinschedulerFeatures\ security features, focusing on user and credential management to authenticate with external services.
\dolphinschedulerLLocPct\% of its code is security feature-related.

\parhead{Apache Hadoop (Data Store)} is a distributed data store and analytics platform for processing large datasets\,\cite{hadoop}.
It offers \hadoopFeatures\ security features in \hadoopLLocPct\% of its code, from access control lists and Kerberos authentication to its own key management system for encryption keys and providing encryption for data at rest using encryption zones and in transit using transport encryption.

\parhead{Apache Hudi (Big Data)} is a data lake platform for managing large datasets\,\cite{hudi}.
Hudi, does not implement security features, but uses \hudiFeatures\ when interacting with other systems, such as Hadoop or cloud storage services.
This requires \hudiLLocPct\% of its code.

\parhead{Apache Iceberg (Big Data)} is a table format for datasets\,\cite{iceberg}.
Similar to Hudi, Iceberg relies on underlying storage and processing systems, such as Hadoop, to enforce access control.
Still, it provides \icebergFeatures\ security features in \icebergLLocPct\% of its code, such as encryption of data.

\parhead{Apache RocketMQ (Streaming and Event System)}, formerly maintained by Alibaba before its donation to the Apache Software Foundation, is a distributed messaging and streaming platform\,\cite{rocketmq}.
RocketMQ uses transport layer security for encrypting data in transit, and uses access control lists to restrict access to topics and queues.
It provides \rocketmqFeatures\ security features in \rocketmqLLocPct\% of its code.

\parhead{Apache Tomcat (Application Server)} is a web server and servlet container for Java applications\,\cite{tomcat}.
Tomcat implements \tomcatFeatures\ security features across \tomcatLLocPct\% of its code, including session management, role-based access control to restrict access to resources based on user or application roles and permissions, and transport encryption to secure data in transit.

\parhead{Eureka (Cloud)} is a microservice management platform for service discovery and registration\,\cite{eureka}.
Leases are used to establish trust between services, ensuring that only authorized services can register and discover other services.
\eurekaLLocPct\% of its code is dedicated to \eurekaFeatures\ security features.

\parhead{Signal Server (Enterprise Service)} is the backend server of the Signal messenger application for secure messaging as well as voice and video calls\,\cite{signal}.
It supports end-to-end encryption for messages and calls and uses a large range of cryptographic features. 
While cryptography is a core security feature, it implements \signalserverFeatures\ security features across \signalserverLLocPct\% of its code.

\parhead{Traccar (IoT Management System)} is a GPS tracking platform for tracking vehicles, phones, and other GPS devices\,\cite{traccar}.
It uses \traccarLLocPct\% of its code for \traccarFeatures\ security features, such as access control, implemented by associating users with specific devices and restricting access to those devices.

%% file: 05_Results.tex
\input{05_ResultsRQ1.tex}
\input{05_ResultsRQ2.tex}
\input{05_ResultsRQ3.tex}
\input{05_ResultsRQ4.tex}

%% file: 05_ResultsRQ1.tex
\section{Footprint of Security Features (RQ1)}
We first quantify the size and scattering of security features across the codebase.
We then qualitatively analyze the factors that contribute to the footprint for individual security features.

\subsection{Quantitative Analysis}
We measure the size of security features in logical lines of code (LLOC) by removing comments and blank lines, and their scattering as the number of files and methods which contain the feature.

\subsubsection{Size of Security Features}

\begin{figure}[t]
\centering
\includegraphics[width=.87\linewidth]{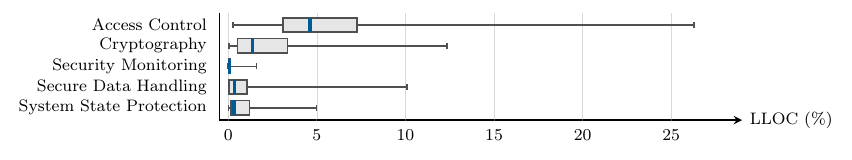}
\vspace{-16pt}
\caption{Relative share of security feature LLOC in relation to respective project LLOC across all projects}
\label{fig:lloc-stats}
\vspace{-18pt}
\end{figure}

\Cref{fig:lloc-stats} shows the share of security feature code in relation to the total LLOC of the respective project.
All projects contain at least one security feature from each of the five top-level categories, except for system state protection, which is absent from RocketMQ. 
Apache Hudi has the smallest share of security feature code. 
In contrast, the Signal server has the largest share of access control, secure data handling, and system state protection, while cryptography and security monitoring have their largest shares in Tomcat.

On average, when comparing security feature sizes relative to the respective project size, access control is the most prevalent security feature (\accesscontrolLLocPctAverage\% $\pm$ \accesscontrolLLocPctStd\%),
followed by cryptography (\cryptographyLLocPctAverage\% $\pm$ \cryptographyLLocPctStd\%).
Secure data handling (\securedatahandlingLLocPctAverage\% $\pm$ \securedatahandlingLLocPctStd\%) and system state protection (\systemstateprotectionLLocPctAverage\% $\pm$ \systemstateprotectionLLocPctStd\%) are less prevalent.
Security monitoring (\securitymonitoringLLocPctAverage\% $\pm$ \securitymonitoringLLocPctStd\%) has the lowest code share of all security features.

Across all security features, the share of code varies considerably between projects, as reflected by high standard deviations relative to the means. As expected, the absolute size of security features strongly correlates with overall project size (Pearson correlation coefficient $r$ = \rqOneAbsSizeR{}), while the relative share of security feature code only weakly correlates negatively with project size ($r$ = \rqOneRelSizeR{}).
Differences in the relative share instead appear to reflect the underlying security model of a project.
Projects that implement comprehensive security models for security-critical infrastructure contain particularly large shares of security feature code, including the Signal server (\signalserverLLocPct\%), the Tomcat server (\tomcatLLocPct\%), and Hadoop (\hadoopLLocPct\%).
In contrast, projects which assume to be deployed in a trusted environment, such as Apache Hudi (\hudiLLocPct\%) and Iceberg (\icebergLLocPct\%), tend to have a smaller share.

\begin{figure}[t]
\centering
\includegraphics[width=.8\linewidth]{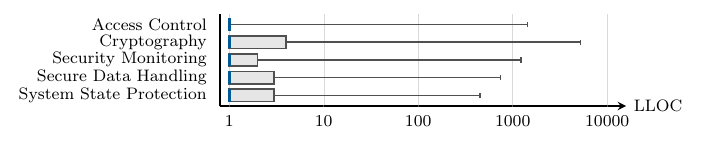}
\vspace{-17pt}
\caption{Size of security feature locations in Logical Lines of Code}
\label{fig:location-size-stats}
\vspace{-14pt}
\end{figure}

\Cref{fig:location-size-stats} complements the relative share of security features with the sizes of their feature locations in LLOC, i.e., contiguous blocks of code implementing a specific security feature.
Most locations are small, with a median of one LLOC across all categories, but their maximum sizes vary considerably.
Cryptography contains the largest individual locations (up to \cryptographyLocSizeMax{} LLOC), corresponding to a single OpenSSL implementation file in Tomcat.
This is followed by access control (up to \accesscontrolLocSizeMax{} LLOC) in in Hadoop's UserGroupInformation, which contains user authentication data, and security monitorng (up to \securitymonitoringLocSizeMax{} LLOC) in Tomcat's AbstractAccessLogValve for HTTP access logging.
The largest locations for secure data handling (\securedatahandlingLocSizeMax{} LLOC) and system state protection (\systemstateprotectionLocSizeMax{} LLOC) are smaller and occur in Signal server's ArchiveController for secure backup storage and Hadoop's LoadBalancingKMSClientProvider for key-management-system load balancing, respectively. 
Overall, access control and cryptography are not only among the largest security feature categories but also contain the largest contiguous blocks of security-related code.

Although access control is the largest security feature overall, its locations are confined to single lines that fetch authentication data or perform authorization checks.
Cryptography, in contrast, has a larger upper quartile (\cryptographyLocSizeQThree{} LLOC), followed closely by secure data handling (\securedatahandlingLocSizeQThree{} LLOC) and system state protection (\systemstateprotectionLocSizeQThree{} LLOC). Security monitoring has the smallest upper quartile (\securitymonitoringLocSizeQThree{} LLOC) after access control, reflecting that many of its locations consist of individual logging statements.

\subsubsection{Scattering}

\begin{table}[b]
\centering
\vspace{-12pt}
\caption{Absolute counts and relative share of security features across project files and methods}
\vspace{-10pt}
\label{tab:scattering-stats}
\smaller
\setlength{\tabcolsep}{2.5pt}
\begin{tabular}{llrrrrr}
\toprule
 & \textbf{Security Feature} & \textbf{Min} & \textbf{Median} & \textbf{Max} & \textbf{Avg} & \textbf{Std} \\
\midrule
\multirow{5}{*}{\rotatebox[origin=c]{90}{File-level}} & Access Control          & \accesscontrolScatMin{} (\accesscontrolScatPctMin\%) & \accesscontrolScatMedian{} (\accesscontrolScatPctMedian\%) & \accesscontrolScatMax{} (\accesscontrolScatPctMax\%) & \accesscontrolScatAverage{} (\accesscontrolScatPctAvg\%) & \accesscontrolScatStd{} (\accesscontrolScatPctStd\%) \\
                     & Cryptography            & \cryptographyScatMin{} (\cryptographyScatPctMin\%) & \cryptographyScatMedian{} (\cryptographyScatPctMedian\%) & \cryptographyScatMax{} (\cryptographyScatPctMax\%) & \cryptographyScatAverage{} (\cryptographyScatPctAvg\%) & \cryptographyScatStd{} (\cryptographyScatPctStd\%) \\
                     & Security Monitoring     & \securitymonitoringScatMin{} (\securitymonitoringScatPctMin\%) & \securitymonitoringScatMedian{} (\securitymonitoringScatPctMedian\%) & \securitymonitoringScatMax{} (\securitymonitoringScatPctMax\%) & \securitymonitoringScatAverage{} (\securitymonitoringScatPctAvg\%) & \securitymonitoringScatStd{} (\securitymonitoringScatPctStd\%) \\
                     & Secure Data Handling    & \securedatahandlingScatMin{} (\securedatahandlingScatPctMin\%) & \securedatahandlingScatMedian{} (\securedatahandlingScatPctMedian\%) & \securedatahandlingScatMax{} (\securedatahandlingScatPctMax\%) & \securedatahandlingScatAverage{} (\securedatahandlingScatPctAvg\%) & \securedatahandlingScatStd{} (\securedatahandlingScatPctStd\%) \\
                     & System State Protection & \systemstateprotectionScatMin{} (\systemstateprotectionScatPctMin\%) & \systemstateprotectionScatMedian{} (\systemstateprotectionScatPctMedian\%) & \systemstateprotectionScatMax{} (\systemstateprotectionScatPctMax\%) & \systemstateprotectionScatAverage{} (\systemstateprotectionScatPctAvg\%) & \systemstateprotectionScatStd{} (\systemstateprotectionScatPctStd\%) \\
\midrule
\multirow{5}{*}{\rotatebox[origin=c]{90}{Method-level}} & Access Control          & \accesscontrolMScatMin{} (\accesscontrolMScatPctMin\%) & \accesscontrolMScatMedian{} (\accesscontrolMScatPctMedian\%) & \accesscontrolMScatMax{} (\accesscontrolMScatPctMax\%) & \accesscontrolMScatAverage{} (\accesscontrolMScatPctAvg\%) & \accesscontrolMScatStd{} (\accesscontrolMScatPctStd\%) \\
                       & Cryptography            & \cryptographyMScatMin{} (\cryptographyMScatPctMin\%) & \cryptographyMScatMedian{} (\cryptographyMScatPctMedian\%) & \cryptographyMScatMax{} (\cryptographyMScatPctMax\%) & \cryptographyMScatAverage{} (\cryptographyMScatPctAvg\%) & \cryptographyMScatStd{} (\cryptographyMScatPctStd\%) \\
                       & Security Monitoring     & \securitymonitoringMScatMin{} (\securitymonitoringMScatPctMin\%) & \securitymonitoringMScatMedian{} (\securitymonitoringMScatPctMedian\%) & \securitymonitoringMScatMax{} (\securitymonitoringMScatPctMax\%) & \securitymonitoringMScatAverage{} (\securitymonitoringMScatPctAvg\%) & \securitymonitoringMScatStd{} (\securitymonitoringMScatPctStd\%) \\
                       & Secure Data Handling    & \securedatahandlingMScatMin{} (\securedatahandlingMScatPctMin\%) & \securedatahandlingMScatMedian{} (\securedatahandlingMScatPctMedian\%) & \securedatahandlingMScatMax{} (\securedatahandlingMScatPctMax\%) & \securedatahandlingMScatAverage{} (\securedatahandlingMScatPctAvg\%) & \securedatahandlingMScatStd{} (\securedatahandlingMScatPctStd\%) \\
                       & System State Protection & \systemstateprotectionMScatMin{} (\systemstateprotectionMScatPctMin\%) & \systemstateprotectionMScatMedian{} (\systemstateprotectionMScatPctMedian\%) & \systemstateprotectionMScatMax{} (\systemstateprotectionMScatPctMax\%) & \systemstateprotectionMScatAverage{} (\systemstateprotectionMScatPctAvg\%) & \systemstateprotectionMScatStd{} (\systemstateprotectionMScatPctStd\%) \\
\bottomrule
\end{tabular}
\vspace{-10pt}
\end{table}

We investigated scattering of security features by counting in how many files and methods they occur, as shown in \cref{tab:scattering-stats}.
We also normalized these counts by the total number of files and methods in each project to characterize their relative distribution.

Similar to security feature size, security features are least scattered in Hudi and most in Tomcat.
An exception is cryptography, which is, despite having less code, most scattered in Signal.

The overall scattering of security features correlates with the absolute project size (r = \rqOneAbsScatFilesR{} for files, r = \rqOneAbsScatMethodsR{} for methods).
The relative scattering of security features, however, like the relative size of security features, is only weakly correlated with project size (r = \rqOneRelScatFilesR{} for files, r = \rqOneRelScatMethodsR{} for methods).
This indicates that while larger projects tend to have larger security features, their relative distribution across the codebase is largely independent of project size.

Access control features are scattered the most across files (\accesscontrolScatPctAvg\% $\pm$ \accesscontrolScatPctStd\%) and methods (\accesscontrolMScatPctAvg\% $\pm$ \accesscontrolMScatPctStd\%).
This is particularly dominated by code for managing user authentication, permissions, and credentials.
Cryptography is the second most scattered security feature across files (\cryptographyScatPctAvg\% $\pm$ \cryptographyScatPctStd\%) and methods (\cryptographyMScatPctAvg\% $\pm$ \cryptographyMScatPctStd\%), heavily influenced by the use of transport layer security and cryptographic key management.
Secure data handling (\securedatahandlingScatPctAvg\% $\pm$ \securedatahandlingScatPctStd\% files and \securedatahandlingMScatPctAvg\% $\pm$ \securedatahandlingMScatPctStd\% methods) and system state protection (\systemstateprotectionScatPctAvg\% $\pm$ \systemstateprotectionScatPctStd\% files and \systemstateprotectionMScatPctAvg\% $\pm$ \systemstateprotectionMScatPctStd\% methods) follow in terms of scattering.
Respectively, secure storage of sensitive data and input validation contribute the most to the scattering of secure data handling and system state protection.
Security monitoring is the least scattered security feature across both files (\securitymonitoringScatPctAvg\% $\pm$ \securitymonitoringScatPctStd\%) and methods (\securitymonitoringMScatPctAvg\% $\pm$ \securitymonitoringMScatPctStd\%).

\subsection{Qualitative Analysis}

We now qualitatively investigate the size and scattering of security features.
\begin{lstlisting}[float,language=Java, caption={Core delegation-token issuer interface in Hadoop (\texttt{DelegationTokenIssuer})}, label={lst:delegation-token-issuer}, belowskip=-16pt]
@InterfaceAudience.LimitedPrivate({"HDFS", "MapReduce", "Yarn"})
public interface DelegationTokenIssuer {
	
	Token<?> getDelegationToken(String renewer) throws IOException; // Request a new delegation token
	
	default Token<?>[] addDelegationTokens(final String renewer, Credentials credentials) {
		...
		collectDelegationTokens(this, renewer, credentials, tokens); // Collect tokens from all existing issuers
		return tokens.toArray(new Token<?>[tokens.size()]);
	}
	
	static void collectDelegationTokens(...) throws IOException { // Recursively gather all tokens for a path
		...
		Token<?> token = credentials.getToken(service);
		token = issuer.getDelegationToken(renewer); 
		credentials.addToken(service, token);
		...
	}
}
\end{lstlisting}
Access control drives much of the size and scattering of security features, encompassing authentication and authorization.
Its implementation is often domain-specific because concepts such as roles, permissions, group memberships, and credentials vary across software systems.
Even when established security mechanisms are used, substantial project-specific access-control code may be required. 
For example, Hadoop uses Kerberos for authentication, but implements additional mechanisms for issuing and collecting delegation tokens, as illustrated by the DelegationTokenIssuer interface in \cref{lst:delegation-token-issuer}. 
Beyond delegation tokens, Hadoop implements further access-control functionality for managing users, access-control lists, authentication tokens, and permissions.

\begin{lstlisting}[float,language=Java, caption={Custom security audit logging for authentication (RocketMQ \texttt{DefaultAuthenticationProvider})}, label={lst:rocketmq-auth-logging}, belowskip=-29pt]
public class DefaultAuthenticationProvider implements AuthenticationProvider<DefaultAuthenticationContext> {
	
	public CompletableFuture<Void> authenticate(DefaultAuthenticationContext ctx) {
		return newHandlerChain().handle(ctx).whenComplete((nil,ex)->doAuditLog(ctx, ex));
	}
	// Audit log for authentication outcomes
	protected void doAuditLog(DefaultAuthenticationContext ctx, Throwable ex) {
		...
		if (ex != null) {
			log.info("[AUTH] User:{} ... failed with Signature = {}.", ctx.getUsername(), ctx.getSignature());
		} else {
			log.debug("[AUTH] User:{} ... success with Signature = {}.", ctx.getUsername(), ctx.getSignature());
		}
	}
}
\end{lstlisting}
%
%
%
%
%

While encryption contributes the most to size and scattering of cryptography, it rarely involves code of cryptographic ciphers.
Instead, much of encryption code concerns transport encryption, certificates, and configuration of secure communication channels.
Integrating concrete block ciphers contributes comparatively little code, whereas the surrounding cryptographic infrastructure, including key generation, management, and revocation, accounts for a larger part.
Cryptographic hashing, however, also contributes notably to the size and scattering of cryptography.

Secure data handling is primarily driven by the secure storage of sensitive data, such as certificates and secrets, and by input validation and sanitization distributed across data-processing paths. Other contributors include retention controls for sensitive data, trusted sources of randomness for cryptographic operations, and parameterized prepared statements for preventing SQL injection. While these mechanisms may occur throughout a codebase, their individual locations are often small and consist of localized checks, configuration, or API usage.

System state protection preserves system integrity in the presence of faults or malicious activity. Its scattering is partly driven by security exceptions, which handle security-relevant erroneous or unexpected conditions throughout the codebase. Session management further contributes through functionality for tracking, authenticating, and invalidating user sessions. Resource management features, such as load balancing and rate limiting, also occur across different parts of a system to control the use of resources and protect against resource exhaustion and denial-of-service attacks.

\quotebox{Footprint of Security Features}{Security features vary considerably in size and scattering across projects. Their absolute footprint increases with project size, while their relative footprint depends more strongly on the security features and security model of the project. Access control has the largest and most scattered footprint, driven by project-specific authentication and authorization logic. Cryptography and secure data handling also contribute substantially, whereas security monitoring has a smaller, but widely distributed footprint, often interacting with access control and cryptography.}{1}

%% file: 05_ResultsRQ2.tex
\section{Security Feature Implementation and Usage Patterns (RQ2)}

\begin{figure}[t]
\centering
\includegraphics[width=\linewidth]{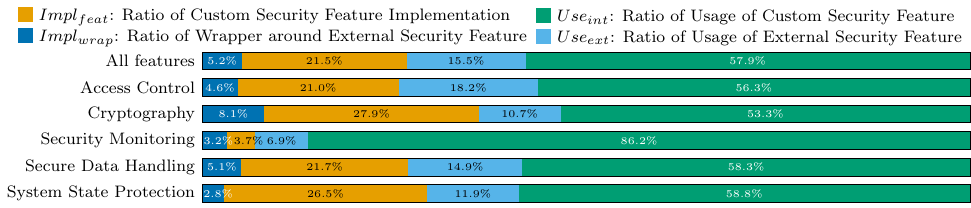}
\vspace{-22pt}
\caption{Percentage of security feature locations following each implementation pattern.}
\label{fig:rq2-patterns}
\vspace{-16pt}
\end{figure}


We identified four implementation and usage patterns.
We distinguish usage ($Use$) from implementation ($Impl$) by whether a location spans a whole Java definition and manually classify external library imports as security-specific or general-purpose.
We assigned each location to a pattern and counted their occurrences.
\Cref{fig:rq2-patterns} shows the distribution of each pattern for each feature.

\parhead{Usage of External Security Feature ($Use_{ext}$).} The location uses a security functionality provided by an external security library.

\parhead{Wrapper around External Security Feature ($Impl_{wrap}$).} The location implements a method that wraps a single security functionality from one namespace of an external security library.

\parhead{Custom Security Feature Implementation ($Impl_{feat}$).} The location implements a security feature by combining multiple externally provided security functionalities or without relying on external security functionality.

\parhead{Usage of Custom Security Feature ($Use_{int}$).} The location uses an internally provided custom implementation of a security feature.

\subsection{Quantitative Analysis}

We first analyze the overall distribution of security feature implementation patterns across the \patUsesTotal\ security feature locations.
We automatically assigned each location to a pattern and counted their occurrences.
Across all security features, usage dominates over implementation, comprising \patCustUsagePct\% use of custom features ($Use_{int}$) and \patExtUsagePct\% use of external security functionality ($Use_{ext}$). 
Custom implementations ($Impl_{feat}$) account for \patCustImplPct\% of locations, while \patExtWrapperPct\% wrap external security functionality ($Impl_{wrap}$).

As the largest and most scattered security feature, access control contributes strongly to the overall distribution of patterns ($Impl_{feat}$: \acPatCustImplPct\%, $Use_{ext}$: \acPatExtUsagePct\%, $Use_{int}$: \acPatCustPct\%, and $Impl_{wrap}$: \acPatExtWrapperPct\%).
Cryptography and system state protection have the highest shares of custom implementations ($Impl_{feat}$: \cryptoPatCustImplPct\% and $Impl_{feat}$: \sspPatCustImplPct\%) and correspondingly low usage of external features ($Use_{ext}$: \cryptoPatExtUsagePct\% and $Use_{ext}$: \sspPatExtUsagePct\%).
Their custom implementations are instead frequently used ($Use_{int}$: \cryptoPatCustUsagePct\%, and $Use_{int}$: \sspPatCustUsagePct\%).
Although wrappers are rare overall, they are most prevalent in cryptography ($Impl_{wrap}$: \cryptoPatExtWrapperPct\%).
Security monitoring relies almost entirely on custom features implemented elsewhere in the project ($Use_{int}$: \monPatCustPct\%), with minimal reliance on external security libraries ($Use_{ext}$: \monPatLibPct\%) or custom implementation ($Impl_{feat}$: \monPatCustImplPct\%).
In contrast, access control and secure data handling have the highest shares of external usage ($Use_{ext}$: \acPatExtUsagePct\% and $Use_{ext}$: \sdhPatExtUsagePct\%).

\begin{figure}[b]
\centering
\vspace{-20pt}
\includegraphics[width=.75\linewidth]{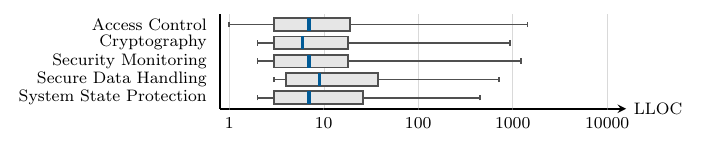}
\vspace{-18pt}
\caption{Size of Custom Implementation per security feature}
\label{fig:rq2-custom-impl-sizes}
\end{figure}

We further investigated the size of custom implementations to estimate how much security-specific code developers write (\cref{fig:rq2-custom-impl-sizes}).
Access control spans the widest range, from one-line methods to classes of up to \accesscontrolCustImplSizeMax\ LLOC, followed by cryptography (\cryptographyCustImplSizeMax\ LLOC) and security monitoring (\securitymonitoringCustImplSizeMax\ LLOC).
Secure data handling and system state protection have smaller maxima (\securedatahandlingCustImplSizeMax\ and \systemstateprotectionCustImplSizeMax\ LLOC).
Despite these extremes, median sizes are low across all security features (between \cryptographyCustImplSizeMedian\ and \securedatahandlingCustImplSizeMedian\ LLOC), showing that custom implementations are typically small. 
Many consist of individual methods, such as getters and setters for credentials or permissions, embedded in larger classes.

\subsection{Qualitative Analysis}

The quantitative analysis shows substantial custom implementation for some security features, motivating a closer examination of its relation to external security functionality.

Access control often combines custom implementations with external security functionality, creating project-specific authentication and authorization abstractions used throughout the codebase.
Credentials illustrate this pattern particularly well: Hadoop implements delegation tokens, Iceberg AWS role assumptions, and DolphinScheduler access tokens on top of externally provided security functionality. 
In Hadoop, the common module \texttt{org.apache.hadoop.security.token.*} contains 35 abstractions for managing delegation tokens, which are used across 418 files in 28 modules for tasks such as file-system access authentication, job submission, and token lifecycle management. 
\Cref{lst:delegation-token-issuer} shows the \texttt{DelegationTokenIssuer} interface, which Hadoop services such as HDFS, YARN, and MapReduce implement to issue and collect delegation tokens.

Cryptography also relies heavily on custom implementations, but these rarely implement cryptographic algorithms from scratch. 
Instead, they primarily integrate existing cryptographic libraries and transport-security mechanisms into project-specific functionality. 
For example, Tomcat contains an OpenSSL package implementing cryptographic functionality across 29 files. 
\Cref{lst:iceberg-ciphers} illustrates how Iceberg uses the JDK's \texttt{AES/GCM/NoPadding} cipher through \texttt{javax.crypto.Cipher} while adding project-specific framing logic for nonces, AAD, and GCM tags. 
Similarly, Hadoop uses a data-holder class \texttt{CipherOption} for negotiating cipher parameters, including the cipher suite, keys, and initialization vectors, which is reused throughout cryptographic handshaking code.

\begin{lstlisting}[float,language=Java, caption={Using an external AES-GCM cipher (Iceberg \texttt{Ciphers})}, label={lst:iceberg-ciphers}, belowskip=-21pt]
public class Ciphers {	
	public static class AesGcmEncryptor {
	
		public AesGcmEncryptor(byte[] keyBytes) {
			this.aesKey = newKey(keyBytes); // Build AES secret key from raw bytes
			this.cipher = newCipher(); // External JDK cipher instance
			this.randomGenerator = new SecureRandom(); // Secure nonce source
			this.nonce = new byte[NONCE_LENGTH]; // 12-byte IV buffer
		}
		
		public int encrypt(byte[] plaintext, ..., byte[] ciphertextBuffer, int ciphertextOffset, byte[] aad) {
			randomGenerator.nextBytes(nonce); // Generate a fresh nonce per encryption
			
			GCMParameterSpec spec = new GCMParameterSpec(GCM_TAG_LENGTH_BITS, nonce);
			cipher.init(Cipher.ENCRYPT_MODE, aesKey, spec); // Initialize AES-GCM
			cipher.updateAAD(aad); // Add authenticated associated data
			// Perform authenticated encryption
			int enciphered = cipher.doFinal(plaintext, ..., ciphertextBuffer, ciphertextOffset + NONCE_LENGTH);
			
			System.arraycopy(nonce, 0, ciphertextBuffer, ciphertextOffset, NONCE_LENGTH);
			return enciphered + NONCE_LENGTH;
		}
	}
		
	private static Cipher newCipher() { return Cipher.getInstance("AES/GCM/NoPadding"); }
}
\end{lstlisting}

Security monitoring relies almost exclusively on custom features, using general-purpose libraries such as SLF4J to record security events of other security features.
For example, to record access attempts or the success or failure of cryptographic operations.
\Cref{lst:rocketmq-auth-logging} shows how RocketMQ augments authentication with audit logging that records the username and verification signature.

\begin{lstlisting}[float,language=Java, caption={Custom secure storage client (Signal Server \texttt{SecureStorageClient})}, label={lst:secure-storage-client}, belowskip=-2pt]
public class SecureStorageClient {	
	
	public SecureStorageClient(final ExternalServiceCredentialsGenerator credentialsGenerator, final Executor executor, final ScheduledExecutorService retryExecutor, final SecureStorageServiceConfiguration configuration) throws CertificateException {
			
		this.credentialsGenerator = credentialsGenerator; // External service credential generator
		this.deleteUri = URI.create(configuration.uri()).resolve(DELETE_PATH); // Target endpoint for deletion
		this.httpClient = FaultTolerantHttpClient.newBuilder() // For HTTPS/TLS transport
			...
			.withSecurityProtocol(FaultTolerantHttpClient.SECURITY_PROTOCOL_TLS_1_3) // Enforce TLS 1.3
			.withTrustedServerCertificates(
				configuration.storageCaCertificates().toArray(new String[0])) // Pin storage root CAs
			.build();
	}
}
\end{lstlisting}

Custom implementations of secure data handling are largely driven by secure storage of sensitive data such as credentials and TLS certificates.
These implementations are subsequently used to store and retrieve sensitive data.
For example, Signal Server implements a custom client for authenticated deletion requests to its secure storage service, pins the service's root certificates, and enforces TLS 1.3 (\cref{lst:secure-storage-client}). 
Signal Server also implements custom data handling for transaction validation and privacy features such as anonymous operations and user discoverability.

For system state protection, much of the external security functionality consists of security exceptions for handling error conditions, whereas mechanisms such as rate limiting and session management are primarily implemented as custom features in Hadoop and Tomcat.

\quotebox{Security Feature Implementation and Usage Patterns}{We identified four patterns in how security features are implemented and used: (1) use of external security features, (2) wrappers around external security features, (3) custom security feature implementations, and (4) use of custom security features. Overall, security feature code is dominated by usage, particularly the reuse of custom implementations throughout the codebase. Custom implementations frequently compose externally provided security functionality into project-specific security features that are subsequently reused throughout the codebase. Cryptography and system state protection have the largest shares of custom implementations, while access control and secure data handling rely most on external security functionality. Security monitoring relies almost exclusively on custom features implemented within the project.}{1}

%% file: 05_ResultsRQ3.tex
\section{Internal and External Security Features (RQ3)}
We characterize security feature origins in terms of internal and external security functionality.

\subsection{Quantitative Analysis}

\begin{figure}[b]
	\vspace{-10pt}
\centering
\includegraphics[width=\linewidth]{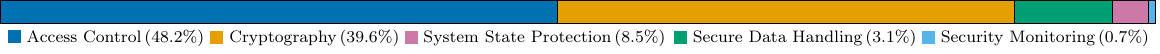}
\vspace{-22pt}
\caption{Share of all identified security functionalities by top-level security feature}
\label{fig:rq3-feature-distribution}
\vspace{-13pt}
\end{figure}

We group Java definitions providing the same semantic security functionality into one distinct security functionality, primarily consolidating overloaded methods.
For example, \texttt{SecureRandom} counts as one distinct security functionality across all its overloaded API methods and usages.
\Cref{fig:rq3-feature-distribution} shows the distribution of the \featureDistTotalSecFns{} distinct security functionalities identified across the five top-level security features.
Access control and cryptography contain the greatest diversity of distinct security functionalities (\featureDistAccessControlSecFnsPct\% and \featureDistCryptographySecFnsPct\%, respectively), followed by secure data handling (\featureDistSecureDataHandlingSecFnsPct\%), system state protection (\featureDistSystemStateProtectionSecFnsPct\%), and security monitoring (\featureDistSecurityMonitoringSecFnsPct\%).

\begin{figure}[t]
	\centering
	\includegraphics[width=.8\linewidth]{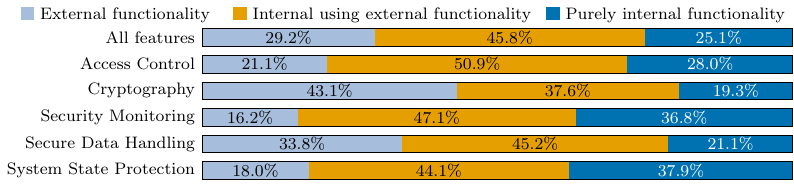}
	\vspace{-10pt}
	\caption{Origin of security-relevant functionality}
	\label{fig:rq3-functionality-origin}
	\vspace{-14pt}
\end{figure}

\Cref{fig:rq3-functionality-origin} distinguishes security functionalities by their origin: external functionality provided by third-party libraries or the JDK, internal implementations relying on external functionality, and purely internal implementations. Of the \secfnsTotal{} identified security functionalities, \rqThreeOriginExtPct\% are external, \rqThreeOriginIntExtPct\% are internal with external dependencies, and \rqThreeOriginPureIntPct\% are purely internal.

Access control is dominated by internal implementations using external functionality (\acOriginIntExtPct\%), but also has a notable purely internal share (\acOriginPureIntPct\%).
Cryptography relies most strongly on external functionality (\cryptoOriginExtPct\%), reflecting the widespread use of JDK and third-party cryptographic APIs.
Security monitoring and system state protection have the largest internal shares (\monOriginPureIntPct\% and \sspOriginPureIntPct\%, respectively), although more than half of their internal implementations use external APIs. 
Secure data handling is more evenly distributed across all three categories.

Overall, most security functionalities rely on external functionality, while purely internal implementations are a minority.
Thus, even custom security code commonly builds on existing external functionality rather than implementing all security functionality internally.

\begin{table}[b]
\centering
\caption{External namespaces containing security APIs covering \rqThreeExtApiTopTwentyOnePct\% of all external security-relevant usages}
\label{tab:rq3-external-apis}
\scriptsize
\addtolength{\tabcolsep}{-2pt}
\vspace{-10pt}
\begin{minipage}[t]{0.345\linewidth}
\begin{tabular}{lrrr}
\toprule
\textbf{Namespace} & \textbf{\makecell[r]{\#F}} & \textbf{\makecell[r]{\#U}} & \textbf{\#P} \\
\midrule
\rqThreeExtApiRootOne & \rqThreeExtApiFnsOne & \rqThreeExtApiUsesOne & \rqThreeExtApiProjOne \\
\rqThreeExtApiRootTwo & \rqThreeExtApiFnsTwo & \rqThreeExtApiUsesTwo & \rqThreeExtApiProjTwo \\
\rqThreeExtApiRootThree & \rqThreeExtApiFnsThree & \rqThreeExtApiUsesThree & \rqThreeExtApiProjThree \\
\rqThreeExtApiRootFour & \rqThreeExtApiFnsFour & \rqThreeExtApiUsesFour & \rqThreeExtApiProjFour \\
\rqThreeExtApiRootFive & \rqThreeExtApiFnsFive & \rqThreeExtApiUsesFive & \rqThreeExtApiProjFive \\
\rqThreeExtApiRootSix & \rqThreeExtApiFnsSix & \rqThreeExtApiUsesSix & \rqThreeExtApiProjSix \\
\rqThreeExtApiRootSeven & \rqThreeExtApiFnsSeven & \rqThreeExtApiUsesSeven & \rqThreeExtApiProjSeven \\
\bottomrule
\end{tabular}
\end{minipage}%
~
\begin{minipage}[t]{0.3\linewidth}
	\begin{tabular}{lrrr}
		\toprule
		\textbf{Namespace} & \textbf{\makecell[r]{\#F}} & \textbf{\makecell[r]{\#U}} & \textbf{\#P} \\
		\midrule
		\rqThreeExtApiRootEight & \rqThreeExtApiFnsEight & \rqThreeExtApiUsesEight & \rqThreeExtApiProjEight \\
		\rqThreeExtApiRootNine & \rqThreeExtApiFnsNine & \rqThreeExtApiUsesNine & \rqThreeExtApiProjNine \\
		\rqThreeExtApiRootTen & \rqThreeExtApiFnsTen & \rqThreeExtApiUsesTen & \rqThreeExtApiProjTen \\
		\rqThreeExtApiRootEleven & \rqThreeExtApiFnsEleven & \rqThreeExtApiUsesEleven & \rqThreeExtApiProjEleven \\
		\rqThreeExtApiRootTwelve & \rqThreeExtApiFnsTwelve & \rqThreeExtApiUsesTwelve & \rqThreeExtApiProjTwelve \\
		\rqThreeExtApiRootThirteen & \rqThreeExtApiFnsThirteen & \rqThreeExtApiUsesThirteen & \rqThreeExtApiProjThirteen \\
		\rqThreeExtApiRootFourteen & \rqThreeExtApiFnsFourteen & \rqThreeExtApiUsesFourteen & \rqThreeExtApiProjFourteen \\
		\bottomrule
	\end{tabular}
\end{minipage}
~
\begin{minipage}[t]{0.28\linewidth}
\begin{tabular}{lrrr}
\toprule
\textbf{Namespace} & \textbf{\makecell[r]{\#F}} & \textbf{\makecell[r]{\#U}} & \textbf{\#P} \\
\midrule
\rqThreeExtApiRootFifteen & \rqThreeExtApiFnsFifteen & \rqThreeExtApiUsesFifteen & \rqThreeExtApiProjFifteen \\
\rqThreeExtApiRootSixteen & \rqThreeExtApiFnsSixteen & \rqThreeExtApiUsesSixteen & \rqThreeExtApiProjSixteen \\
\rqThreeExtApiRootSeventeen & \rqThreeExtApiFnsSeventeen & \rqThreeExtApiUsesSeventeen & \rqThreeExtApiProjSeventeen \\
\rqThreeExtApiRootEighteen & \rqThreeExtApiFnsEighteen & \rqThreeExtApiUsesEighteen & \rqThreeExtApiProjEighteen \\
\rqThreeExtApiRootNineteen & \rqThreeExtApiFnsNineteen & \rqThreeExtApiUsesNineteen & \rqThreeExtApiProjNineteen \\
\rqThreeExtApiRootTwenty & \rqThreeExtApiFnsTwenty & \rqThreeExtApiUsesTwenty & \rqThreeExtApiProjTwenty \\
\rqThreeExtApiRootTwentyOne & \rqThreeExtApiFnsTwentyOne & \rqThreeExtApiUsesTwentyOne & \rqThreeExtApiProjTwentyOne \\
\bottomrule
\end{tabular}
\end{minipage}
\smallskip

\centering
{\tiny \#F: functionalities, \#U: usages, \#P projects}
\vspace{-16pt}
\end{table}

\Cref{tab:rq3-external-apis} shows the 21 most frequently used external security API roots.
Of \rqThreeExtApiTotalNamespaces\ distinct namespace roots, these account for \rqThreeExtApiTopTwentyOneUses\ of \rqThreeExtApiTotalUses\ external uses (\rqThreeExtApiTopTwentyOnePct\%).
The three most prevalent JDK security namespaces, \texttt{java.security}, \texttt{javax.net}, and \texttt{javax.security}, together account for \rqThreeJdkSecUses\ uses (\rqThreeJdkSecUsesPct\%).
Third-party libraries are more project-specific, including \texttt{software.amazon.awssdk} (five projects), \texttt{org.apache.zookeeper} (Hadoop and RocketMQ), and \texttt{io.dropwizard.auth} (Signal).

\subsection{Qualitative Analysis}
The JDK provides substantial security functionality. The extent of internally provided functionality varies between projects and security features.
Infrastructure projects such as Hadoop, Tomcat, and RocketMQ provide more security functionality internally, whereas projects such as DolphinScheduler and Hudi rely more on external APIs.
Security features within the top-level categories exhibit varying degrees of reliance on internal versus external functionality and often combine both.

At the level of individual security features, the balance between internal and external functionality varies substantially.
Credentials commonly combine both, with 67.5\% of custom security functionalities relying on external APIs such as \texttt{java.security.Principal}, \texttt{javax.security.auth.Subject}, and \texttt{software.amazon.awssdk.auth.credentials}.
Some security features, in contrast, are realized almost entirely through external APIs.
Parameterized prepared statements rely entirely on \texttt{java.sql.PreparedStatement}, while secure randomness almost exclusively uses \texttt{java.securi\-ty.SecureRandom} (99.4\%) and concrete block ciphers commonly use \texttt{javax.crypto.Cipher} (93.1\%).
The same pattern occurs for message authentication (100\%), certificate authentication (100\%), and group key management (93.3\%).
Among more prevalent security features, certification also relies predominantly on external APIs (64.1\%), particularly \texttt{java.securi\-ty.cert.X509Certificate}, while sharing APIs with secure communication.
Hadoop provides an exception by introducing its own abstractions for cryptographic operations in \texttt{org.apache.hadoop.crypto}.

Conversely, purely internal implementations remain common (\rqThreeOriginPureIntPct\%), particularly for security features tied to project-specific state and policies.
Session timeout (76.7\%), session management (58.4\%), and role-based access control (55.6\%) are predominantly implemented internally.
Traccar, Tomcat, and DolphinScheduler, for example, provide custom abstractions for sessions, cookies, and session configuration, while Tomcat, Hadoop, and Iceberg define their own abstractions for application roles, user roles, and AWS role assumptions.

\quotebox{Internal and External Security Features}{Most security functionality is implemented within the projects, but the majority of these internal implementations relies on external security functionality. The balance varies substantially between security features: credentials commonly combine internal and external functionality, standardized mechanisms such as secure randomness and block ciphers are predominantly external, while session management and role-based access control contain large shares of purely internal functionality. External functionality is concentrated in few APIs, with 21 namespace roots accounting for \rqThreeExtApiTopTwentyOnePct\% of all external uses and JDK security APIs contributing the largest share.}{1}


%% file: 05_ResultsRQ4.tex
\section{Interaction of Security Features (RQ4)}
We analyze interactions between security features based on their co-occurrence within the code.

\subsection{Quantitative Analysis}

\begin{table}[t]
		\centering
		\caption{Tangling between top-level security features}
		\label{fig:tangling-matrices}
		\vspace{-10pt}
		\setlength{\tabcolsep}{1pt}%
		\newcommand{\tanglingSymText}{PineGreen!35!black}%
		\resizebox{\textwidth}{!}{%
		\begin{tabular}{r|rrrrrcrrrrr|r}
			\toprule
			\makecell[c]{Security\\Features} & \makecell[c]{System State\\Protection} & \makecell[c]{Secure Data\\Handling} & \makecell[c]{Security\\Monitoring} & \makecell[c]{Crypto-\\graphy} & \makecell[c]{Access\\Control} &
			\makecell[r]{%
		\begin{tikzpicture}[baseline=0pt]
			\node[anchor=south east, font=\scriptsize, inner sep=0pt] (lbl) at (-0.15, 0.32) {code contained in};
			\draw[-stealth, semithick, rounded corners=1pt] (-0.15, -0.1) |- (-0.9, 0.20);
			\node[anchor=south west, font=\scriptsize, inner sep=0pt] (lbl) at (0.15, 0.32) {code contains};
			\draw[-stealth, semithick, rounded corners=1pt] (0.15, -0.1) |- (0.9, 0.20);
		\end{tikzpicture}%
			} & \makecell[c]{Access\\Control}& \makecell[c]{Crypto-\\graphy} & \makecell[c]{Security\\Monitoring} & \makecell[c]{Secure Data\\Handling} & \makecell[c]{System State\\Protection} & \makecell[c]{Security\\Features} \\
			\midrule
			\cellcolor{cellSymACSF}\textcolor{\tanglingSymText}{\tanglingSymACtoSF\%} & \cellcolor{cellSymACSS}\textcolor{\tanglingSymText}{\tanglingSymACtoSS\%} & \cellcolor{cellSymACSD}\textcolor{\tanglingSymText}{\tanglingSymACtoSD\%} & \cellcolor{cellSymACSM}\textcolor{\tanglingSymText}{\tanglingSymACtoSM\%} & \cellcolor{cellSymACCR}\textcolor{\tanglingSymText}{\tanglingSymACtoCR\%} & \cellcolor{cellSymACAC}\textcolor{\tanglingSymText}{\tanglingSymACtoAC\%} & Access Control          & \cellcolor{cellACAC}\textcolor{NavyBlue!30!black}{\tanglingACtoAC\%} & \cellcolor{cellACCR}\textcolor{NavyBlue!30!black}{\tanglingACtoCR\%} & \cellcolor{cellACSM}\textcolor{NavyBlue!30!black}{\tanglingACtoSM\%} & \cellcolor{cellACSD}\textcolor{NavyBlue!30!black}{\tanglingACtoSD\%} & \cellcolor{cellACSS}\textcolor{NavyBlue!30!black}{\tanglingACtoSS\%} & \cellcolor{cellACSF}\textcolor{white}{\tanglingACtoSF\%} \\
			\cellcolor{cellSymCRSF}\textcolor{\tanglingSymText}{\tanglingSymCRtoSF\%} & \cellcolor{cellSymCRSS}\textcolor{\tanglingSymText}{\tanglingSymCRtoSS\%} & \cellcolor{cellSymCRSD}\textcolor{\tanglingSymText}{\tanglingSymCRtoSD\%} & \cellcolor{cellSymCRSM}\textcolor{\tanglingSymText}{\tanglingSymCRtoSM\%} & \cellcolor{cellSymCRCR}\textcolor{\tanglingSymText}{\tanglingSymCRtoCR\%} & \cellcolor{cellSymCRAC}\textcolor{\tanglingSymText}{\tanglingSymCRtoAC\%} & Cryptography            & \cellcolor{cellCRAC}\textcolor{NavyBlue!30!black}{\tanglingCRtoAC\%} & \cellcolor{cellCRCR}\textcolor{NavyBlue!30!black}{\tanglingCRtoCR\%} & \cellcolor{cellCRSM}\textcolor{NavyBlue!30!black}{\tanglingCRtoSM\%} & \cellcolor{cellCRSD}\textcolor{NavyBlue!30!black}{\tanglingCRtoSD\%} & \cellcolor{cellCRSS}\textcolor{NavyBlue!30!black}{\tanglingCRtoSS\%} & \cellcolor{cellCRSF}\textcolor{white}{\tanglingCRtoSF\%} \\
			\cellcolor{cellSymSMSF}\textcolor{white}{\tanglingSymSMtoSF\%} & \cellcolor{cellSymSMSS}\textcolor{\tanglingSymText}{\tanglingSymSMtoSS\%} & \cellcolor{cellSymSMSD}\textcolor{\tanglingSymText}{\tanglingSymSMtoSD\%} & \cellcolor{cellSymSMSM}\textcolor{\tanglingSymText}{\tanglingSymSMtoSM\%} & \cellcolor{cellSymSMCR}\textcolor{\tanglingSymText}{\tanglingSymSMtoCR\%} & \cellcolor{cellSymSMAC}\textcolor{white}{\tanglingSymSMtoAC\%} & Security Monitoring     & \cellcolor{cellSMAC}\textcolor{NavyBlue!30!black}{\tanglingSMtoAC\%} & \cellcolor{cellSMCR}\textcolor{NavyBlue!30!black}{\tanglingSMtoCR\%} & \cellcolor{cellSMSM}\textcolor{NavyBlue!30!black}{\tanglingSMtoSM\%} & \cellcolor{cellSMSD}\textcolor{NavyBlue!30!black}{\tanglingSMtoSD\%} & \cellcolor{cellSMSS}\textcolor{NavyBlue!30!black}{\tanglingSMtoSS\%} & \cellcolor{cellSMSF}\textcolor{NavyBlue!30!black}{\tanglingSMtoSF\%} \\
			\cellcolor{cellSymSDSF}\textcolor{white}{\tanglingSymSDtoSF\%} & \cellcolor{cellSymSDSS}\textcolor{\tanglingSymText}{\tanglingSymSDtoSS\%} & \cellcolor{cellSymSDSD}\textcolor{\tanglingSymText}{\tanglingSymSDtoSD\%} & \cellcolor{cellSymSDSM}\textcolor{\tanglingSymText}{\tanglingSymSDtoSM\%} & \cellcolor{cellSymSDCR}\textcolor{\tanglingSymText}{\tanglingSymSDtoCR\%} & \cellcolor{cellSymSDAC}\textcolor{\tanglingSymText}{\tanglingSymSDtoAC\%} & Secure Data Handling    & \cellcolor{cellSDAC}\textcolor{white}{\tanglingSDtoAC\%} & \cellcolor{cellSDCR}\textcolor{NavyBlue!30!black}{\tanglingSDtoCR\%} & \cellcolor{cellSDSM}\textcolor{NavyBlue!30!black}{\tanglingSDtoSM\%} & \cellcolor{cellSDSD}\textcolor{NavyBlue!30!black}{\tanglingSDtoSD\%} & \cellcolor{cellSDSS}\textcolor{NavyBlue!30!black}{\tanglingSDtoSS\%} & \cellcolor{cellSDSF}\textcolor{white}{\tanglingSDtoSF\%} \\
			\cellcolor{cellSymSSSF}\textcolor{\tanglingSymText}{\tanglingSymSStoSF\%} & \cellcolor{cellSymSSSS}\textcolor{\tanglingSymText}{\tanglingSymSStoSS\%} & \cellcolor{cellSymSSSD}\textcolor{\tanglingSymText}{\tanglingSymSStoSD\%} & \cellcolor{cellSymSSSM}\textcolor{\tanglingSymText}{\tanglingSymSStoSM\%} & \cellcolor{cellSymSSCR}\textcolor{\tanglingSymText}{\tanglingSymSStoCR\%} & \cellcolor{cellSymSSAC}\textcolor{\tanglingSymText}{\tanglingSymSStoAC\%} & System State Protection & \cellcolor{cellSSAC}\textcolor{NavyBlue!30!black}{\tanglingSStoAC\%} & \cellcolor{cellSSCR}\textcolor{NavyBlue!30!black}{\tanglingSStoCR\%} & \cellcolor{cellSSSM}\textcolor{NavyBlue!30!black}{\tanglingSStoSM\%} & \cellcolor{cellSSSD}\textcolor{NavyBlue!30!black}{\tanglingSStoSD\%} & \cellcolor{cellSSSS}\textcolor{NavyBlue!30!black}{\tanglingSStoSS\%} & \cellcolor{cellSSSF}\textcolor{NavyBlue!30!black}{\tanglingSStoSF\%} \\
			\midrule
			\cellcolor{cellSymSFSF}\textcolor{\tanglingSymText}{\tanglingSymSFtoSF\%} & \cellcolor{cellSymSFSS}\textcolor{\tanglingSymText}{\tanglingSymSFtoSS\%} & \cellcolor{cellSymSFSD}\textcolor{\tanglingSymText}{\tanglingSymSFtoSD\%} & \cellcolor{cellSymSFSM}\textcolor{\tanglingSymText}{\tanglingSymSFtoSM\%} & \cellcolor{cellSymSFCR}\textcolor{\tanglingSymText}{\tanglingSymSFtoCR\%} & \cellcolor{cellSymSFAC}\textcolor{\tanglingSymText}{\tanglingSymSFtoAC\%} & Security Features & \cellcolor{cellSFAC}\textcolor{NavyBlue!30!black}{\tanglingSFtoAC\%} & \cellcolor{cellSFCR}\textcolor{NavyBlue!30!black}{\tanglingSFtoCR\%} & \cellcolor{cellSFSM}\textcolor{NavyBlue!30!black}{\tanglingSFtoSM\%} & \cellcolor{cellSFSD}\textcolor{NavyBlue!30!black}{\tanglingSFtoSD\%} & \cellcolor{cellSFSS}\textcolor{NavyBlue!30!black}{\tanglingSFtoSS\%} & \cellcolor{cellSFSF}\textcolor{white}{\tanglingSFtoSF\%} \\
			\bottomrule
		\end{tabular}%
		}
		\vspace{-18pt}
	\end{table}

We quantify interactions by measuring how much of one security feature's code overlaps with another.
As overlap is directional, the interaction matrix is not necessarily symmetric.
\Cref{fig:tangling-matrices} reports two perspectives: the share of each feature's code that contains another feature and the share contained within another feature. Within the same top-level category, we consider only interactions between distinct security features (e.g., credentials do not interact with themselves).

Overall, \tanglingSymSFtoSF\% of security feature code overlaps with another security feature.
Secure data handling contains the most code of other features (\tanglingSDtoSF\%), particularly access control (\tanglingSDtoAC\%) and cryptography (\tanglingSDtoCR\%).
Access control follows (\tanglingACtoSF\%), primarily interacting with other access control features (\tanglingACtoAC\%) and cryptography (\tanglingACtoCR\%).
Cryptography (\tanglingCRtoSF\%) interacts most strongly within its own category (\tanglingCRtoCR\%), reflecting interactions among features such as key management, encryption, and hashing. Security monitoring (\tanglingSMtoSF\%) and system state protection (\tanglingSStoSF\%) also contain other security features, for example when access-control events are logged (\tanglingSMtoAC\%).

From the opposite perspective, most security monitoring code is contained within other security features (\tanglingSymSMtoSF\%), particularly access control (\tanglingSymSMtoAC\%) and cryptography (\tanglingSymSMtoCR\%), reflecting the logging of access control events and cryptographic operations.
Secure data handling is also frequently contained within other features (\tanglingSymSDtoSF\%), particularly access control (\tanglingSymSDtoAC\%).
Cryptography is predominantly contained within itself (\tanglingSymCRtoCR\%), followed by access control (\tanglingSymCRtoAC\%).
Access control (\tanglingSymACtoSF\%) and system state protection (\tanglingSymSStoSF\%) are less frequently contained within other features.

Overall, interactions most frequently involve access control (\tanglingSFtoAC\%) and cryptography (\tanglingSFtoCR\%). 
Interactions occur between all category pairs, while within-category interactions occur for all categories except security monitoring.
Because security monitoring and system state protection are comparatively small, their interactions also account for relatively little shared code.

\subsection{Qualitative Analysis}

We examine how the observed interactions manifest in code and why particular security features occur together.
Access control, the largest security feature, participates in many interactions.
Within access control, authentication establishes identities and credentials that authorization and role-based access control use for access decisions.
Credentials require cryptographic protection and secure storage, connecting access control to cryptography and secure data handling.
Hadoop's delegation tokens illustrate the interaction between credentials and token issuing (\cref{lst:delegation-token-issuer}).

These dependencies also explain interactions across security feature categories. 
Secure data handling connects access control and cryptography, as sensitive data requires both restricted access and cryptographic protection. 
Security monitoring, in turn, is embedded in other security features to record security-relevant events and the outcomes of security operations. 
For example, \cref{lst:rocketmq-auth-logging} shows how RocketMQ directly couples authentication with security audit logging.

The quantitative analysis revealed directional differences in interactions, with access control more often contained within secure data handling than vice versa.
We observed that secure data operations are often guarded by access control checks.
In Signal, for example, \texttt{deleteEntireBackup} invokes \texttt{checkBackupLevel} before deleting a user's backup.
Similarly, security monitoring is commonly embedded in access control operations rather than the reverse.

Interactions also occur between features within the same category.
Cryptography shows the strongest such interactions, as key management, key generation, encryption, hashing, and signatures often form part of the same cryptographic workflows.

Finally, the quantitative analysis shows substantial differences in interaction across projects, with Hadoop, Tomcat, and Signal Server exhibiting more interaction than Hudi, Iceberg, and Eureka.
Our inspection of these differences points to the projects' security models as an important factor, as they shape which security features are required and how they interact.

\quotebox{Interaction of Security Features}{Security features interact by depending on and being embedded within one another, with \tanglingSymSFtoSF\% of security feature code co-occurring with other security features. Access control interacts with cryptography and secure data handling through credentials and sensitive data, while cryptographic features depend on one another for operations such as encryption and key management. Security monitoring is typically embedded within access control and cryptography to record security-relevant operations. The extent and structure of these interactions vary across projects and appear related to their underlying security models.}{1}

%% file: 06_Discussion.tex
\section{Related Work}

Prior research on software security has predominantly focused on identifying vulnerabilities rather than characterizing the implementation of security functionality in software systems.
For example, Henriques et al.\,\cite{Henriques2024} mine vulnerability databases and source-code repositories to continuously update vulnerability datasets, while Shihab and Alanezi\,\cite{Shihab2024} survey datasets and
techniques used for software vulnerability analysis.
Çetin et al.\,\cite{Cetin2024} investigate the use of large language models for detecting vulnerabilities in source code.
These works concern the presence and detection of security defects. 
In contrast, we study the code that realizes \emph{functional security features}, code that mitigates security risks, rather than posing them.

More closely related to our objective, prior work has investigated the identification and interpretation of security-relevant code.
Ryoo et al.\,\cite{ryoo2016use} study the use of architectural security tactics in open source software projects.
Mirakhorli et al.\,\cite{mirakhorli2016detecting} use machine learning to identify security tactics.
Similarly, SWAN\,\cite{piskachev2019codebase} uses machine learning to detect security-relevant methods for sanitization and authentication, while Dev-Assist\,\cite{Johnson2024} builds on it to automatically generate specifications for static analysis.
Krishnamurthy et al.\,\cite{krishnamurthy2026detection} use large language models to explain security-critical code with the goal of helping developers avoid introducing vulnerabilities.
These approaches aim to locate or reason about selected security-relevant architecture decisions in code, whereas we characterize how functional security features are realized across complete software systems.


Overall, existing work provides substantial evidence on vulnerabilities and security code analysis, but offers little empirical evidence on how security features are realized.
Our work addresses this question by studying security features in nine Java systems and characterizing their size, scattering, tangling, implementation patterns, and reliance on internal and
external functionality.

\section{Discussion}

Our findings show that security features extend beyond individual security API calls.
Their implementation crosses library, location, and feature boundaries, with substantial variation across projects.
These characteristics have implications for security feature location, security analysis, and security-aware development tools.

\parhead{Security Is More Than a Library Call.}
Security libraries are important building blocks of security features, but are only a part of them.
While their APIs can provide useful entry points for locating security features\,\cite{hermann2025taxonomy}, our results show that these entry points often lead to project-specific abstractions that are used throughout the codebase.
Security feature location techniques should treat APIs as seeds and follow them into custom implementations and their usages.

Locating such functionality is generally challenging\,\cite{hermann2025engineering}, and general-purpose feature-location techniques struggle with scalability and false positives\,\cite{rubin2013survey,dit2013feature}.
An alternative to repeatedly recovering feature locations is to maintain feature-to-code knowledge during evolution\,\cite{hermann2025visualizing}.
Our findings suggest that such knowledge should capture external functionality, project-specific implementations, and their usage sites.
For practitioners, existing project-specific security abstractions can provide starting points for understanding and modifying security features.
Our dataset captures these relationships, providing a basis for security-specific feature-location techniques.

\parhead{Security Features Are Scattered and Tangled.}
Our findings connect to a long line of feature-interaction research showing that features cannot always be understood independently\,\cite{apel2013featureinteraction,apel2011featureaware,soares2018featureinteraction}.
Prior work has also identified structural interactions at the source-code level\,\cite{schuster2014structural}.
We observe related interactions for security features in general-purpose software systems: access-control checks occur within secure-data operations, credentials require cryptographic protection, and security monitoring is embedded within access-control and cryptographic functionality.

Scattering distributes security functionality across the codebase, requiring reasoning across multiple implementation locations\,\cite{Melo2016,Passos2021}, while tangling additionally requires reasoning across feature boundaries.
Existing feature-interaction techniques provide a starting point for identifying and reasoning about such dependencies\,\cite{apel2011featureaware,apel2013featureinteraction,atlee2015measuring}.
Our dataset enables investigating how these techniques can be adapted to functional security features in conventional software systems.

These characteristics are also relevant to vulnerability analysis.
Static security analyzers struggle with false positives and practical adoption\,\cite{Johnson2013,Bennett2024,hermann2026toolsexploitablevulnerabilitiessurvey}.
TraceSec\,\cite{peldszus2026} shows that providing project-specific security context can improve the prioritization of analyzer findings.
Security-feature locations and interactions provide complementary code-level context that could help analyses account for related functionality beyond individual APIs or locations.
Future work could investigate whether this context improves vulnerability detection and prioritization.

\parhead{Security Features Depend on Project Context.}
Security-feature footprint and interaction vary substantially across projects and appear to depend more on the underlying security model than on project size. 
Researchers and tools should therefore account for project-specific security responsibilities and abstractions rather than assuming a universal implementation structure.

This variation means that the relevant context for implementing a security feature is inherently project-specific, which is particularly important for AI-assisted development.
Developers already struggle with security APIs\,\cite{nadi2016jumping,acar2016information,acar2017crypto,green2016developers,naiakshina2017password,naiakshina2019if}, and AI-generated code can contain vulnerabilities\,\cite{pearce2025asleep,ambati2024navigating,dincua2024ai,serafini2025exploring,perry2023users}, while coding agents may operate on a limited scope of the codebase\,\cite{horikawa2025agentic}.
Our findings suggest that security-aware agents should consider existing security abstractions, their usage sites, and interacting features when modifying or generating security-sensitive code.
Our dataset provides annotated examples for investigating whether such context improves AI-assisted security development.

%% file: 07_Limitations.tex
\section{Threats to Validity}
We now discuss the potential threats to the validity of our study.

\parhead{Internal Validity.}
Author bias may have influenced the sampling of subject systems.
The systems were selected based on the knowledge and familiarity with the subject domain of two authors, resulting in 9 systems from 7 different categories.
Author bias may also have influenced the identification of security features.
While the first author conducted the identification, the first and second author frequently discussed and reviewed the dataset to mitigate potential biases.
The categorization of security features according to the taxonomy may also be subject to author bias, which we mitigated through discussion between two authors until consensus was reached.

\parhead{External Validity.}
The generalizability of our findings may be limited due to how we selected the subject systems.
Our study focused on software systems written in Java, which may not generalize to systems written in other programming languages.
The subject systems were selected based on their popularity on Github, which are often mature and security-conscious.
Finally, our subject systems were drawn from 7 different categories, which may limit the generalizability of our findings to other types of software systems.

\parhead{Conclusion Validity.}
The correlations reported in our study may be prone to outliers in our dataset.
During sampling, we aimed to select projects that are representative of different categories, and span varying sizes, organizations, and security focus.

\parhead{Construct Validity.}
We used a predefined taxonomy of security features for the annotation and categorization process, which may constrain the scope of security features considered.
The authors discussed all security features that we could not immediately categorize until consensus was reached.
We were able to categorize all security features according to the predefined taxonomy.

%% file: 08_Conclusion.tex
%
%
%

\section{Conclusion}

We presented a mining study of how functional security features are realized in nine open-source Java software systems.
We manually annotated their security-feature code and analyzed its footprint, implementation and usage patterns, origin, and interactions.

Our results show that security features are more than just a simple library call.
While individual code locations are often small, they collectively form a substantial and scattered footprint across the codebase.
Security-feature code is dominated by the use of internally implemented functionality that frequently integrates external security functionality into project-specific abstractions.
Security features also interact with each other, and their implementations often overlap.

Our annotated dataset provides a basis for developing and evaluating techniques for security feature location, security analysis, and security-aware development.
While external security APIs provide useful entry points for locating security features, analyses should also account for project-specific implementations, their usages, and interacting features.
Security functionality therefore needs to be considered across libraries, locations, and feature interactions.